\documentclass[twocolumn]{elsarticle}

\usepackage{lineno}
\modulolinenumbers[5]

\begin{document}

\begin{frontmatter}
\onecolumn

\title{SPring-8 LEPS2 beamline: A facility to produce a multi-GeV photon beam via 
laser Compton scattering}


\author[elph]{N.~Muramatsu}
%\author[elph]{N.~Muramatsu\corref{correspondingauthor}}
%\cortext[correspondingauthor]{Corresponding author}
%\ead{mura@lns.tohoku.ac.jp}

\author[rcnp]{M.~Yosoi}
\author[rcnp]{T.~Yorita}
\author[rcnp]{Y.~Ohashi}

\author[korea]{J.K.~Ahn}
\author[rcnp]{S.~Ajimura}
\author[xfel]{Y.~Asano}
\author[acasin]{W.C.~Chang}
\author[nsrrc]{J.Y.~Chen}
\author[rcnp]{S.~Dat\'{e}}
\author[kyoto]{T.~Gogami}
\author[rcnp]{H.~Hamano}
\author[rcnp]{T.~Hashimoto}
\author[harima]{T.~Hiraiwa}
\author[rcnp]{T.~Hotta}
\author[elph]{T.~Ishikawa}
\author[rcnp]{Y.~Kasamatsu}
\author[rcnp]{H.~Katsuragawa}
\author[rcnp]{R.~Kobayakawa}
\author[rcnp]{H.~Kohri}
\author[tokyo]{S.~Masumoto}
\author[elph]{Y.~Matsumura}
\author[elph]{M.~Miyabe}
\author[jlab]{K.~Mizutani}
\author[kek]{Y.~Morino}
\author[rcnp]{T.~Nakano}
\author[rcnp]{T.~Nam}
\author[kyosan]{M.~Niiyama}
\author[uoth]{Y.~Nozawa}
\author[rcnp,aichi]{H.~Ohkuma}
\author[elph]{H.~Ohnishi}
\author[uoth]{T.~Ohta}
\author[jasri]{M.~Oishi}
\author[kek,tokyo]{K.~Ozawa}
\author[rcnp]{S.Y.~Ryu}
\author[elph]{Y.~Sada}
\author[elph]{H.~Saito}
\author[tokyo]{T.~Shibukawa}
\author[elph]{H.~Shimizu}
\author[elph]{R.~Shirai}
\author[jasri]{M.~Shoji}
\author[gifu,rcnp]{M.~Sumihama}
\author[jasri]{S.~Suzuki}
\author[rcnp]{S.~Tanaka}
\author[jasri]{Y.~Taniuchi}
\author[elph]{A.~O.~Tokiyasu}
\author[rcnp]{N.~Tomida}
\author[jparc]{Y.~Tsuchikawa}
\author[rcnp]{K.~Watanabe}
\author[rcnp]{C.J.~Yoon}
\author[elph]{C.~Yoshida}

\address[elph]{Research Center for Electron Photon Science, Tohoku University, Sendai, 
               Miyagi 982-0826, Japan}
\address[rcnp]{Research Center for Nuclear Physics, Osaka University, Ibaraki, Osaka 
               567-0047, Japan}
\address[korea]{Department of Physics, Korea University, Seoul 02841, Republic of Korea}
\address[xfel]{XFEL Project Head Office, RIKEN, Sayo, Hyogo 679-5143, Japan}
\address[acasin]{Institute of Physics, Academia Sinica, Taipei 11529, Taiwan}
\address[nsrrc]{National Synchrotron Radiation Research Center, Hsinchu 30076, Taiwan}
\address[kyoto]{Department of Physics, Kyoto University, Kyoto 606-8502, Japan}
\address[harima]{RIKEN SPring-8 Center, Sayo, Hyogo 679-5148, Japan}
\address[tokyo]{Department of Physics, University of Tokyo, Tokyo 113-0033, Japan}
\address[jlab]{Thomas Jefferson National Accelerator Facility, Newport News, Virginia 
               23606. USA}
\address[kek]{Institute of Particle and Nuclear Studies, High Energy Accelerator
               Research Organization (KEK), Tsukuba, Ibaraki 305-0801, Japan}
\address[kyosan]{Department of Physics, Kyoto Sangyo University, Kyoto 603-8555, Japan}
\address[uoth]{Department of Radiology, The University of Tokyo Hospital, Tokyo 113-8655, 
               Japan}
\address[aichi]{Aichi Synchrotron Radiation Center, Seto, Aichi 489-0965, Japan}
\address[jasri]{Japan Synchrotron Radiation Research Institute (SPring-8), Sayo, Hyogo 
               679-5198, Japan}
\address[gifu]{Department of Education, Gifu University, Gifu 501-1193, Japan}
\address[jparc]{J-PARC Center, Japan Atomic Energy Agency, Tokai, Ibaraki 319-1195, Japan}

\begin{abstract}
We have constructed a new laser-Compton-scattering facility, called the LEPS2 beamline, 
at the $8$-$\mathrm{GeV}$ electron storage ring, SPring-8. This facility provides 
a linearly polarized photon beam in a tagged energy range of $1.3$--$2.4$~$\mathrm{GeV}$. 
Thanks to a small divergence of the low-emittance storage-ring electrons, the tagged 
photon beam has a size ($\sigma$) suppressed to about $4$~$\mathrm{mm}$ even after it 
travels about $130$~$\mathrm{m}$ to the experimental building that is independent of 
the storage ring building and contains large detector systems. This beamline is designed 
to achieve a photon beam intensity higher than that of the first laser-Compton-scattering 
beamline at SPring-8 by adopting the simultaneous injection of up to four high-power 
laser beams and increasing a transmittance for the long photon-beam path up to about 
$77$\%. The new beamline is under operation for hadron photoproduction experiments.
\end{abstract}

\begin{keyword}
high energy photon beams, laser Compton scattering, linear polarization, hadron photoproduction
%\texttt{elsarticle.cls}\sep \LaTeX\sep Elsevier \sep template
%\MSC[2010] 00-01\sep  99-00
\end{keyword}

\end{frontmatter}
\twocolumn
%\linenumbers

\section{Introduction}

Photon beams in the energy range from a few hundred $\mathrm{MeV}$ to several 
$\mathrm{GeV}$ are useful tools to investigate the structures and properties of hadrons, 
which can be produced by the reaction of a photon with a target nucleon or nucleus (hadron 
photoproduction). In particular, photon beams with a few $\mathrm{GeV}$ energies are 
suitable for hadron studies in the strange sector. As a distinct feature, photons can be 
polarized linearly or circularly. This feature is effective to obtain the spin and parity 
information of hadrons in photoproduction reactions. Currently, high-energy photon beams 
for hadron photoproduction experiments are generated by laser Compton scattering or 
bremsstrahlung at several facilities in the world. So far, facilities based on the laser 
Compton scattering were realized at SPring-8 (LEPS) \cite{nimlepsbl, monograph}, ESRF 
(GRAAL) \cite{graal}, and BNL (LEGS) \cite{legs}. However, only the LEPS2 beamline at 
SPring-8, described in this article, is now under operation as the laser Compton 
scattering facility. On the other hand, the high-energy photon-beam production by 
bremsstrahlung has been achieved at JLab \cite{jlab}, ELSA \cite{elsa}, MAMI \cite{mami}, 
ELPH \cite{elph1, elph2} etc.

For laser Compton scattering, high-intensity laser light is injected into a high-energy 
electron storage ring so that photons should be backscattered over a very narrow angular 
region \cite{feenberg, milburn, arutyunyan}. At this scattering, the energy of a photon 
is greatly amplified by receiving a large fraction of the energy of an electron. While
the beam production by bremsstrahlung requires a dedicated electron accelerator, the laser 
Compton scattering facility has an advantage of low-cost operation at one of many beamlines 
in an electron storage ring. In addition, demands for the electron storage rings are 
increasing as synchrotron radiation sources for material, earth, life sciences etc. 
Therefore, the laser Compton scattering is utilized not only for hadron photoproduction 
experiments in the $\mathrm{GeV}$ energy region but also for various researches and 
applications with $\mathrm{MeV}$ photon beams \cite{photonucl, vortex, nuclsecur}. The 
energy spectrum of a photon beam obtained by laser Compton scattering has a continuously
spread shape with a maximum at the Compton edge, whereas the intensity of bremsstrahlung 
decreases in inverse proportion to the photon energy. The photon beam production by laser 
Compton scattering significantly reduces a low-energy component, which cannot be tagged 
in the energy measurement and contributes as experimental backgrounds.

Laser light is almost $100$\% linearly polarized, and it is easy to control the direction 
of a linear polarization or make circular polarization by using a waveplate (retarder) 
\cite{waveplate}. The polarization state of the laser light can be transferred to the 
photon beam by Compton scattering in a wide energy range. For both linearly and circularly 
polarized photon beams, the degree of polarization increases as the energy goes up to the 
Compton edge \cite{dangelo, lepsproj}. On the other hand, a linearly polarized component
of the bremsstrahlung beam is generated in a relatively narrow energy range (coherent 
bremsstrahlung \cite{cohbrems}), and the polarization decreases as this range is adjusted 
to the higher energy side. A high-energy coherent bremsstrahlung beam with high 
linear-polarization is only achieved by increasing the energy of an electron beam, as 
recently done in JLab for the photon beam energies around $9$~$\mathrm{GeV}$ \cite{gluex}. 
In the energy range from approximately $2$ to several $\mathrm{GeV}$, there are no 
bremsstrahlung facilities providing high linear-polarization for a photon beam. Instead, 
the laser Compton scattering beamlines at SPring-8 have generated a highly 
linear-polarized photon beam in the energy range up to the Compton edge, exceeding 
$2$~$\mathrm{GeV}$. For the production of a circularly polarized photon beam, a highly 
polarized electron beam is necessary at bremsstrahlung facilities \cite{circpol} with 
a high cost. Whereas, the circular polarization of a laser-Compton-scattering beam can 
be obtained only by the optical control of incident laser light.

The LEPS beamline (BL33LEP) has been operated since 1999 as the first laser Compton 
scattering facility at SPring-8, which is a storage ring with an electron energy of 
$7.975$~$\mathrm{GeV}$. The photon beam energy at this beamline is the world's highest 
among the facilities adopting laser Compton scattering. When the energy of incident 
laser light, the energy of an electron storage ring, and the rest mass of an electron 
are denoted by $ k $, $E_e$, and $m_e$, respectively, the maximum energy of a photon 
beam via laser Compton scattering ($E_{\gamma}^\mathrm{max}$) is written as
\begin{linenomath}
\begin{equation}
E_{\gamma}^\mathrm{max} \simeq \frac{4 {E_e}^2 k}{{m_e}^2 c^4 + 4 E_e k} . \label{eqn:maxene}
\end{equation}
\end{linenomath}
At the LEPS beamline, ultraviolet (UV) laser light with a wavelength of $355$~$\mathrm{nm}$ 
($k = 3.49$~$\mathrm{eV}$) is injected into SPring-8 to produce a photon beam in the energy
range up to $E_{\gamma}^\mathrm{max} = 2.39$~$\mathrm{GeV}$ \cite{nimlepsbl}. The intensity 
of this photon beam (the counting rate of photons that are tagged and reach a detector system 
for hadron photoproduction experiments) has been about $10^6$~$\mathrm{s^{-1}}$ with the laser
output power of $8$~$\mathrm{W}$. The stabilization and maximization of a Compton scattering 
rate have been realized by the top-up operation that constantly refills electrons to the 
storage ring up to $100$~$\mathrm{mA}$ \cite{topup}. Nevertheless, the obtained beam 
intensity is still an order of magnitude lower than that of the bremsstrahlung facilities. 
In order to get a higher photon-beam intensity without accelerator upgrades, it is necessary 
to increase the total power of incident laser light and the photon-beam transmittance over
the entire beamline. The incident power can be improved by operating multiple high-power UV 
lasers simultaneously, and the transmittance goes up by the reduction of beamline materials. 

For the photoproduction of heavier hadrons, it is also an important issue to further raise
the photon beam energy while attaining a reasonable beam intensity. Advances in the laser 
technology have made it possible to shorten the wavelength of incident light in Compton 
scattering for the production of a higher-energy photon beam. The LEPS beamline has succeeded 
in generating a photon beam up to the energy $E_{\gamma}^\mathrm{max} = 2.89$~$\mathrm{GeV}$ 
by using deep-UV lasers with a wavelength of $266$~$\mathrm{nm}$ ($k = 4.66$~$\mathrm{eV}$). 
However, the output power of an existing deep-UV laser is about an order of magnitude lower 
than that of the UV laser with a wavelength of $355$~$\mathrm{nm}$. As a result, the photon 
beam intensity achieved by the injection of deep-UV lasers has been limited to about 
$10^5$~$\mathrm{s^{-1}}$. It is indispensable to develop high-power lasers also for the 
deep-UV wavelength and construct a system for the simultaneous injection of them. These
developments in the laser injection is essential to obtain a higher-energy photon beam in
future because the electron beam energies of new storage rings tend to be reduced below the 
SPring-8 energy in order to aim for higher-brilliance synchrotron radiation.

For the purpose to solve the above problems of the photon beam intensity and energy, the 
LEPS2 beamline (BL31LEP) has been newly constructed as the second laser Compton scattering 
facility at SPring-8. Its operation has started in 2013 \cite{baryons2013}. The LEPS2 
beamline is designed to allow the injection of up to four laser beams with either of the UV 
($355$~$\mathrm{nm}$) or deep-UV ($266$~$\mathrm{nm}$) wavelength. New high-power lasers are 
also being introduced step by step. In addition, the LEPS2 beamline has succeeded in taking 
a sufficient space to construct large and complex detector-systems for hadron photoproduction 
experiments. So far, the LEPS beamline has been built in the experimental hall common to many 
other beamlines, and the LEPS experimental setup has been made
compact by specializing in the detection of charged particles that are produced at extremely 
forward angles \cite{lepsproj, lepsdet}. In the case of the LEPS2 beamline, experimental 
setups that can detect both charged and neutral particles over almost all solid angles have 
been assembled in a large space inside the LEPS2 experimental building. This building is 
independent of the storage ring building housing the common experimental hall together. Two 
types of large-acceptance detector systems with different characteristics are installed into 
the LEPS2 experimental building, and alternately operated according to experimental programs. 
The construction of the LEPS2 experimental building was possible because a spread of the
photon beam is well suppressed thanks to a small divergence of the electron beam in the 
LEPS2 beamline.

This article describes the details of the LEPS2 beamline facility together with the design 
concepts. In addition, the method of photon beam production in this facility is given with
discussions about the observed beam properties. The following part is organized by six
sections, covering an overview of the beamline design (Sec.~\ref{sec:design}), laser 
injection and polarization measurement systems (Sec.~\ref{sec:laser}), beamline equipment 
for the photon beam production and transportation (Sec.~\ref{sec:bmline}), the LEPS2 
experimental building containing detector systems for hadron photoproduction experiments 
(Sec.~\ref{sec:expbldg}), a detector system tagging high-energy photons generated by laser 
Compton scattering (Sec.~\ref{sec:tagger}), and the properties of a photon beam in the LEPS2 
beamline (Sec.~\ref{sec:property}). Finally, a summary follows in Sec.~\ref{sec:summary} 
with future prospects.

\section{Beamline design} \label{sec:design}

For the photon beam production by laser Compton scattering, the LEPS2 beamline uses one 
of only four long straight sections in SPring-8. A length of the part without any magnets 
reaches $30$~$\mathrm{m}$ in this straight section. In contrast, the LEPS beamline is
constructed along one of the 40 usual straight sections, which are $7.8$~$\mathrm{m}$ 
long. Electron beam divergence in the straight section influences the angular spread of 
a generated photon beam. The electron beam emittance of SPring-8 has become better 
from 2013 \cite{emittance}, and standard deviations ($\sigma$'s) of the electron beam 
divergence at the long straight section are currently $8$ and $0.7$~$\mathrm{\mu rad}$ 
in the horizontal and vertical directions, respectively. This divergence is much better 
than that of the $7.8$~$\mathrm{m}$ straight section, where the $\sigma$'s are $57$ and 
$1.4$~$\mathrm{\mu rad}$ in the above two directions.

\begin{figure*}[htbp]
  \centering
  \includegraphics[width=14cm,bb=0 0 589 265]{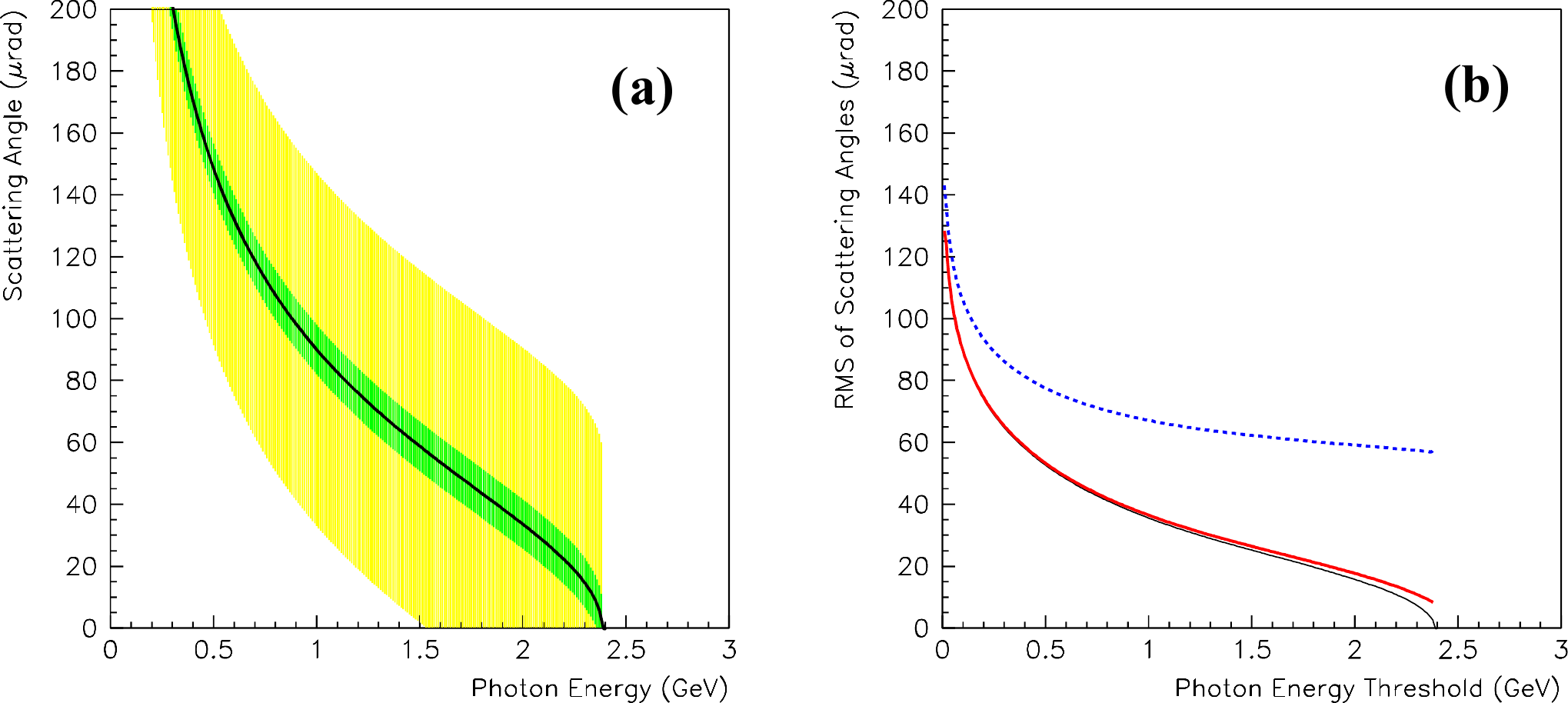}
  \caption{(a) The energy dependence of Compton scattering angles for photons produced 
           in a conical shape (black line). The kinematics was calculated in the case 
           of injecting laser light with a wavelength of $355$~$\mathrm{nm}$ into 
           SPring-8. The green and yellow areas indicate the $1 \sigma$ spreads when 
           the horizontal divergences of the electron beams in the LEPS2 and LEPS 
           beamlines are smeared, respectively. (b) The RMS of scattering angles in the
           horizontal projection plane for the photon-beam energy range exceeding the 
           threshold shown in the x-axis. The conditions of photon beam production are 
           the same as those for (a). The black thin solid, red thick solid, and blue 
           dashed lines show the cases for the kinematics of laser Compton scattering, 
           the LEPS2 beamline, and the LEPS beamline, respectively.}
  \label{fig:beam_divergence}
\end{figure*}
In the head-on collision of laser and electron beams, the correlation between the 
backscattering angle of photons $\theta$ and the amplified energy $E_{\gamma}$ is 
kinematically determined by the following equation:
\begin{linenomath}
\begin{equation}
E_{\gamma} = \frac{k (1 + \beta_e)}{(1 - \beta_e \cos \theta) 
             + \frac{k (1 + \cos \theta)}{E_e}}  \label{eqn:eneang}
\end{equation}
\end{linenomath}
This equation tells that photons generated by laser Compton scattering are emitted 
in a conical shape with a scattering angle depending on the photon energy, as shown in 
Fig.~\ref{fig:beam_divergence}(a). Kinematically, the scattering angle $\theta$ becomes 
$0$ degrees at the maximum energy corresponding to the Compton edge. Actual scattering 
angles are influenced by the electron beam divergence as indicated with the filled areas 
in Fig.~\ref{fig:beam_divergence}(a). These areas show the $1 \sigma$ spreads of 
scattering angles after smearing the horizontal divergences of the electron beams in 
the LEPS2 and LEPS beamlines. In other words, they express the realistic production 
angles of photons that are scattered in the horizontal plane including the electron 
beam axis. There is a large discrepancy between the angular spreads in the two beamlines 
due to different electron-beam divergences.

Figure~\ref{fig:beam_divergence}(b) shows the RMS of scattering angles in the horizontal 
projection plane as a function of the lower bound (or the threshold value) of a selected 
photon-beam energy range. Contributions from the energy range exceeding the threshold are 
integrated with weights according to the photon beam energy distribution. A size of the 
photon beam in the selected energy range is given by multiplying a tangent of this RMS 
value by a distance. The RMS values were obtained by simulating photon scattering 
directions which form an energy-dependent cone and convolving them with the electron 
beam divergence of the LEPS2 or LEPS beamline. The evaluated horizontal angular spread 
for the LEPS2 beamline is not much different from the result using the simple kinematics 
of laser Compton scattering without any convolution. The horizontal divergence of 
a photon beam in the LEPS2 beamline is suppressed to about $30$~$\mathrm{\mu rad}$ for 
the tagged energy range above $1.3$~$\mathrm{GeV}$. In the case of the LEPS beamline, 
the horizontal angular spread is calculated to be $62$~$\mathrm{\mu rad}$ by the same 
procedure except for changing the lower bound of the tagged energy range to 
$1.5$~$\mathrm{GeV}$. The smaller angular spread in the LEPS2 beamline allows the 
transportation of a photon beam to a location far from the Compton scattering point, 
so that the LEPS2 experimental building containing large detector systems has been 
constructed about $130$~$\mathrm{m}$ downstream. The photon beam size in the LEPS2 
experimental building is suppressed to $\sigma \sim 4$~$\mathrm{mm}$ for the tagged
energy range. As suggested by Fig.~\ref{fig:beam_divergence}(b), a size variation of
the photon beam can also be observed depending on the photon energy region in the LEPS2 
beamline, but not in the LEPS beamline.

The LEPS2 beamline is designed to make it possible to inject a maximum of four laser 
beams simultaneously. Each incident laser beam is expanded to a $1/e^2$ diameter of 
about $40$~$\mathrm{mm}$ by a beam expander to focus on a narrow spot corresponding to 
the electron beam size at the Compton scattering point, which is about $31.5$~$\mathrm{m}$ 
away from the laser injection system. The four expanded laser beams are injected into
the storage ring so that their profiles should be in contact with each other by arranging 
the central axes of four incident laser beams at the vertices of a square. Increase of 
the overall incident laser power causes a higher intensity of the photon beam, although 
the intensity gain is not fully proportional to the total power due to laser incident 
angles relative to the electron beam. An ideal gain factor of the photon beam intensity 
by the four laser injection has been simulated as shown in Fig.~9 of Ref.~\cite{nimlepsbl}.
It is evaluated to be $2.14$. Furthermore, we are improving the photon beam intensity 
also by introducing high-power lasers with a wavelength of $355$~$\mathrm{nm}$, which 
is currently chosen for normal operations. So far, such UV lasers with the output powers 
of $16$ and $24$~$\mathrm{W}$ have been installed in addition to $8$~$\mathrm{W}$ lasers,
which have been mainly used in the LEPS beamline. 

The explained injection method using multiple lasers also enables the minimization of 
photon beam loss during its transportation from the Compton scattering point to the 
LEPS2 experimental building. The center of the square whose vertices correspond to the 
four laser beam axes is a peripheral or low power-density part of any laser light.
It is also a region that becomes the path of a photon beam produced by laser Compton 
scattering. Therefore, a hole has been made in the center of a mirror that reflects 
the four laser beams toward the Compton scattering point. This procedure has reduced 
the amount of beamline materials, maximizing the number of photons reaching the LEPS2 
experimental building.

Propagation of the laser light follows Gaussian beam optics. When a designed value 
of the laser beam size is set at a focusing point (or the Compton scattering point), 
the magnification of a beam expander is uniquely determined as a function of the focal 
length. (See Eq.~(2) of Ref.~\cite{nimlepsbl}.) As this length becomes longer, it is 
necessary to increase the magnification factor and correspondingly enlarge the apertures 
of optical devices (e.g. reflection mirrors) and ultra-high vacuum chambers for laser 
beam paths. In order to make the focal length as short as possible in the LEPS2 beamline, 
laser beams are orthogonally injected into a beamline chamber through a hole at a side 
concrete wall of the storage ring tunnel and reflected to the long straight section by 
a mirror in a vacuum. 

Vacuum chambers connected to the long straight section have a special specification that 
cuts the internal diameter larger than the normal aperture for synchrotron radiation. 
Sufficient apertures that four laser beams go through are secured. A bottleneck of the 
laser beam passage exists in the vertical opening of a vacuum chamber inside the bending 
magnet that is $13.5$~$\mathrm{m}$ away from the Compton scattering point. This vertical 
opening is enlarged to a maximum of $\pm 20$~$\mathrm{mm}$, accepting $2.5 \sigma$
regions of the laser beam profiles. The internal aperture of the LEPS2 beamline has 
a margin allowing even elliptical focusing of laser beams with a cylindrical expander 
\cite{nimlepsbl} as a future option. In this case, by doubling the cross section of 
an incident laser beam only in the vertical direction, the vertical beam size at the 
focal point can be halved with twice the power density. Such laser beam shaping will
enhance the efficiency of Compton scattering with the stored electron beam, whose cross 
section is flat having the horizontal and vertical sizes ($\sigma$) of $340$ and 
$10$~$\mathrm{\mu m}$, respectively. 

\begin{figure*}[t]
  \centering
  \includegraphics[width=16cm,bb=0 0 590 432]{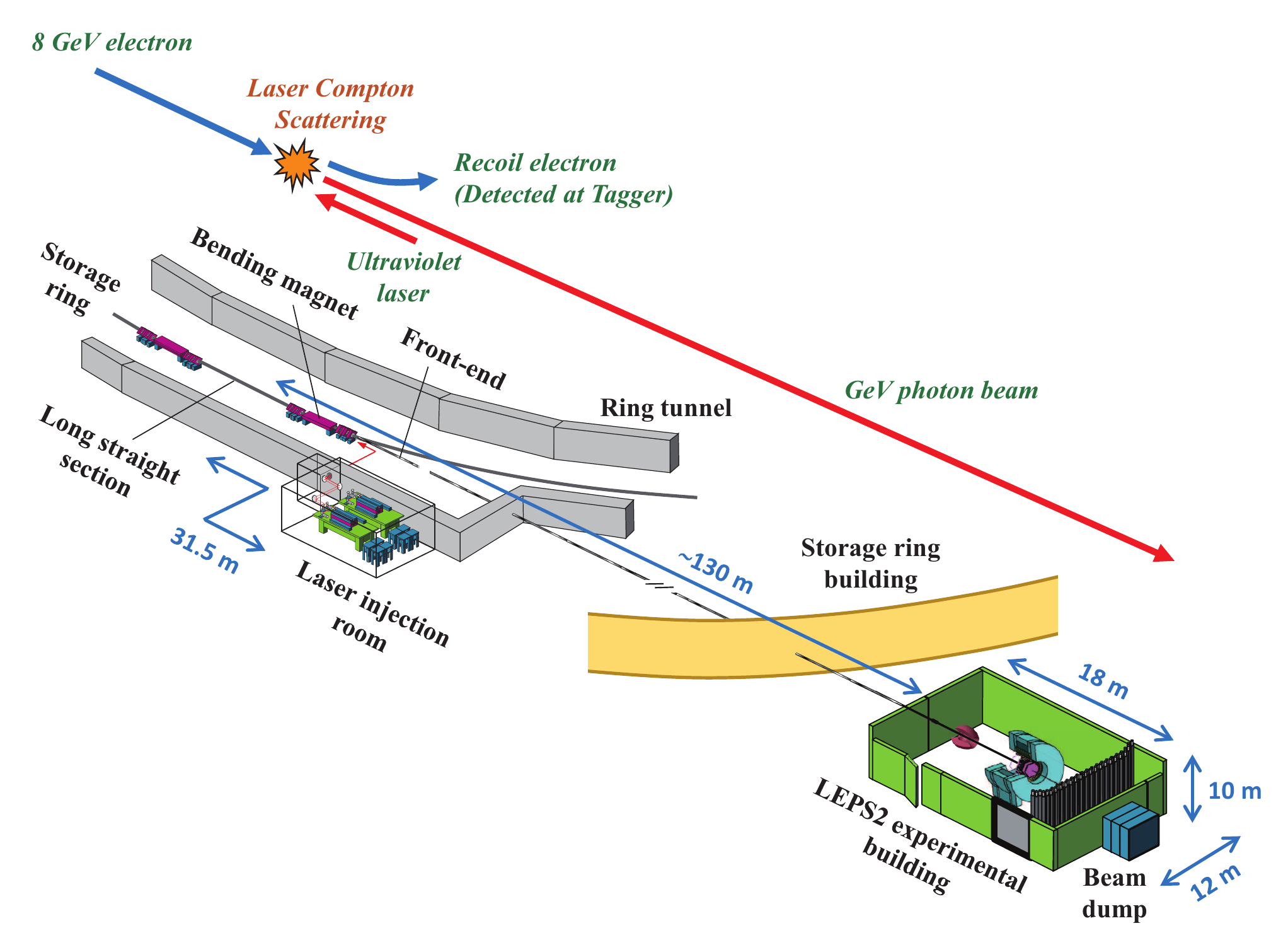}
  \caption{A conceptual overview of the LEPS2 beamline.}
  \label{fig:overview_leps2bl}
\end{figure*}
Figure~\ref{fig:overview_leps2bl} shows an outline drawing of the LEPS2 beamline. Laser 
oscillators and a part of optical devices are arranged on a surface plate in the laser 
injection room with an area of approximately $5$~$\mathrm{m}$ $\times$ $8$~$\mathrm{m}$. 
As described above, multiple laser beams are merged so as to be adjacent to each other 
after the beam sizes are individually magnified by beam expanders. The merged beams are
then guided into the storage ring tunnel by using three large mirrors with a diameter 
of $150$~$\mathrm{mm}$. As a laser beam path, a hole with a diameter of 
$160$~$\mathrm{mm}$ is drilled through the $1$-$\mathrm{m}$ thick concrete wall that 
forms the storage ring tunnel. A radiation protection hatch is constructed only around 
this hole in the experimental hall of the storage ring building. One of the three large 
mirrors mentioned above is placed in the radiation protection hatch. 

The laser beams entering the tunnel are injected into a beamline chamber of the 
``front-end'' section from its side entrance port. Chambers in the front-end section
are directly connected to the storage ring on the extension of the long straight section 
with a ultra-high vacuum, which is typically around $10^{-8}$--$10^{-7}$~$\mathrm{Pa}$. 
Inside the chamber that has the side entrance port, the ``first mirror'' has been
installed to reflect the incident laser beams to the upstream long straight section. 
In addition, the ``monitor mirror'' which drives up and down has been set up in 
a different chamber $0.7$~$\mathrm{m}$ upstream of the first mirror. In the case of 
extracting the incident laser beams to the polarization measurement system, the laser
beam direction can be changed to the outside of the front-end section by the monitor 
mirror. The monitor mirror is retracted upward when a photon beam is generated by 
laser Compton scattering. 

The laser beams pass through the bending magnet, and then travel through the straight 
chambers that are also commonly used as the stored electron beam path. The section of 
those straight chambers is equipped with quadrupole and sextupole magnets for a distance 
of $10$~$\mathrm{m}$. Finally they reach the $30$-$\mathrm{m}$ long straight section 
where no magnets exist. The beamline is designed to cause Compton scattering at 
a position about $2$~$\mathrm{m}$ after the laser beams enter this straight section. 
At a location $8.4$~$\mathrm{m}$ upstream from the Compton scattering point, there is 
a vacuum chamber that monitors laser spots and allows visual confirmation of the laser 
beam axes.

A photon beam generated by laser Compton scattering returns in the opposite direction 
of the incident laser-beam path. Recoil electrons which have lost energies due to 
Compton scattering also follow the same path. Their trajectories are largely bent to 
the inner side of the storage ring at the bending magnet. A detector for recoil 
electrons, called ``tagger'', is installed downstream of the bending magnet to measure 
the hit position of a recoil electron whose trajectory differs depending on its momentum. 
A photon beam, which is a flux of high-energy $\gamma$ rays, goes straight through the 
bending magnet and passes through a hole made at the center of the first mirror without 
loss due to electron-positron pair creation. After exiting a front-end vacuum chamber 
through a $2$-$\mathrm{mm}$ thick aluminum window, it passes through a collimator and 
a sweep magnet. The collimator is located $24.5$~$\mathrm{m}$ downstream from the Compton 
scattering point.

From downstream of the sweep magnet, the photon beam enters a transport pipe with a vacuum 
of about $10$~$\mathrm{Pa}$, and goes out of the storage ring tunnel. This transport pipe 
connects a long distance from the storage ring building to the LEPS2 experimental building, 
which is constructed about $130$~$\mathrm{m}$ downstream from the Compton scattering point. 
The photon beam is extracted into the atmosphere from the transport pipe, and delivered to 
the detector systems that are set up for hadron photoproduction experiments in the LEPS2 
experimental building. The height of the LEPS2 beamline is set to $1200$~$\mathrm{mm}$ 
inside the storage ring tunnel, and $1400$~$\mathrm{mm}$ in all other places.

\section{Laser injection system} \label{sec:laser}

\begin{figure*}[t]
  \centering
  \includegraphics[width=16cm,bb=0 0 258 167]{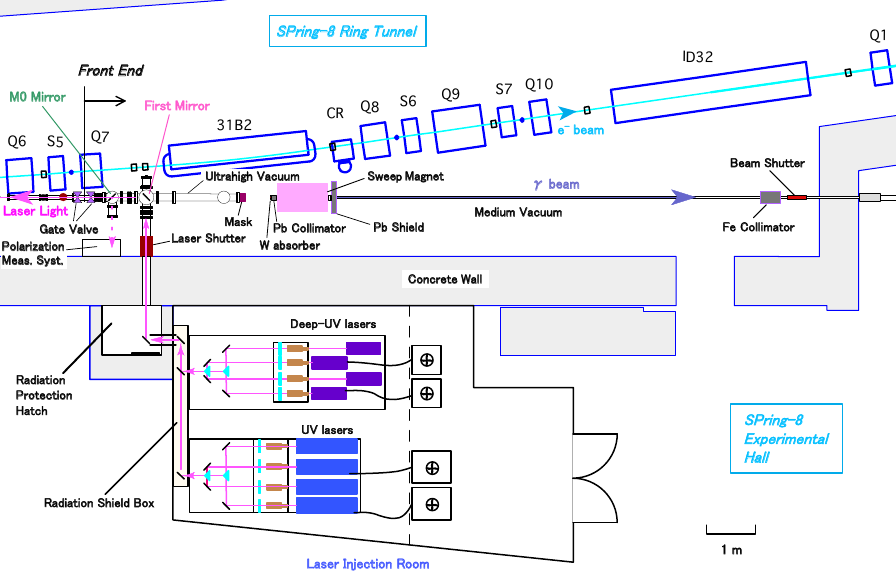}
  \caption{A plan view of the front-end section with the laser injection system
           in the LEPS2 beamline.}
  \label{fig:laser_injection}
\end{figure*}
\begin{figure*}[t]
  \centering
  \includegraphics[width=13cm,bb=0 0 589 445]{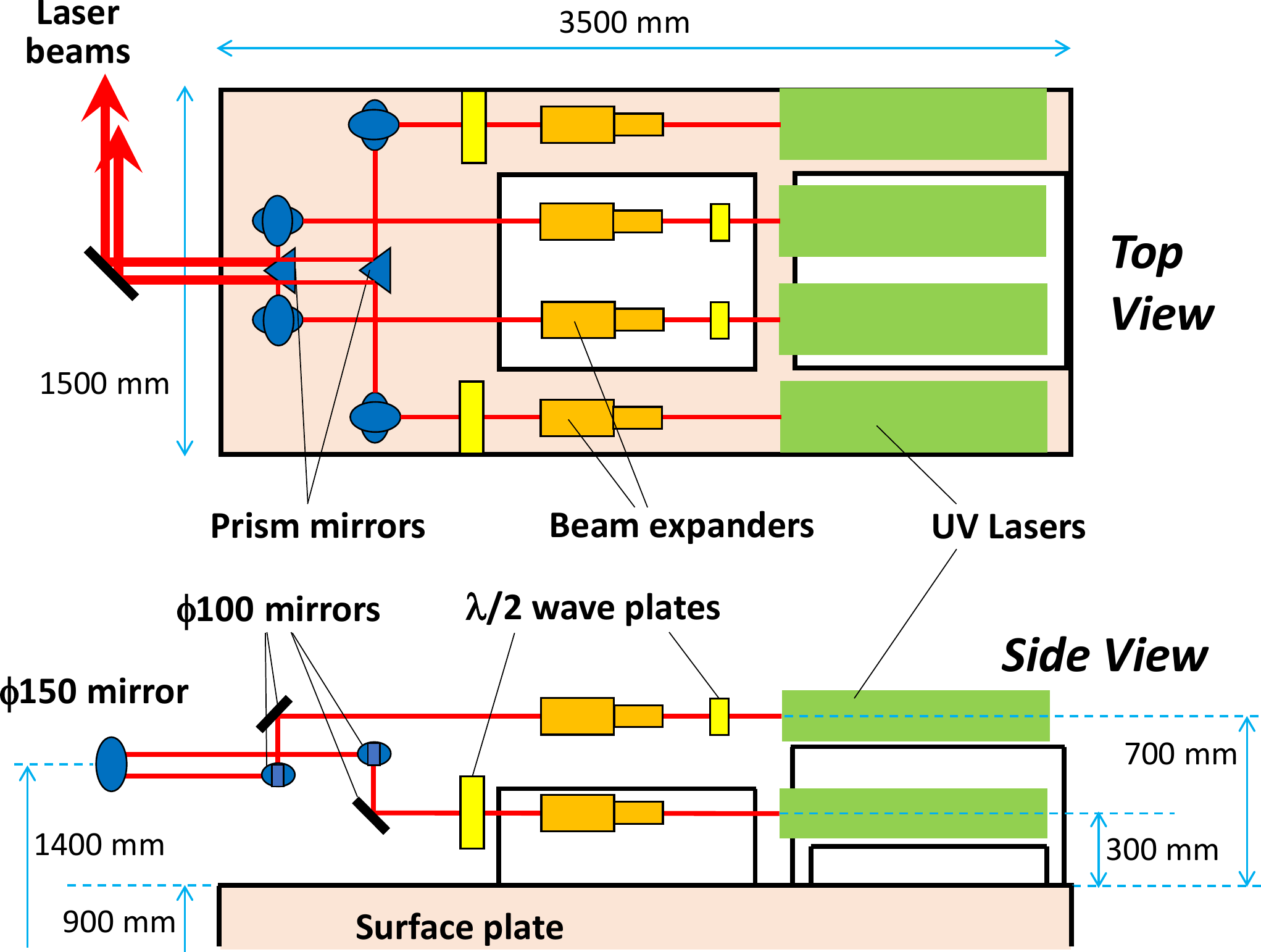}
  \caption{The optical system to simultaneously inject four laser beams into
           SPring-8. All devices are on a large surface plate.}
  \label{fig:laser_room}
\end{figure*}

Figure~\ref{fig:laser_injection} is a plan view of the section related to laser beam
injection in the LEPS2 beamline. The optical system is built on a surface plate in 
an aluminum booth, called the laser injection room. This booth is made as a simple 
clean room under the condition of ISO class 7. We have operated UV lasers 
with a wavelength of $355$~$\mathrm{nm}$ and an output power of $8$, $16$, or 
$24$~$\mathrm{W}$ (Paladin Advanced, manufactured by Coherent Inc.\cite{paladin}). 
Deep-UV lasers with a wavelength of $266$~$\mathrm{nm}$ are also usable on another 
surface plate. As shown in Fig.~\ref{fig:laser_room}, two stages are prepared with 
different heights, and four UV lasers are installed so that two of them should be 
placed on each stage. A laser beam output from the individual oscillator is optically 
controlled by a set of a beam expander, a waveplate, and two mirrors.

The beam expander expands the $1/e^2$ diameter of an incident UV laser beam from 
$1.35$~$\mathrm{mm}$ to about $40$~$\mathrm{mm}$ for the purpose to make a focus at 
the Compton scattering point, which is located $31.5$~$\mathrm{m}$ ahead. On the basis 
of Gaussian beam optics \cite{nimlepsbl, monograph}, the focal length is adjusted by 
changing a magnification factor of the beam expander, where a distance between the 
concave (input) and convex (output) lenses is optimized by a micrometer. This 
adjustment is finally accomplished by maximizing a Compton scattering rate or 
a photon beam intensity, measured by the tagger described in Sec.~\ref{sec:tagger}.

The two mirrors on the surface plate are used to adjust the direction and position 
of a laser beam injected into the storage ring. They are made of a synthetic quartz 
substrate with a diameter of $80$~$\mathrm{mm}$. Their reflective surface is treated 
with a dielectric coating giving a reflectance of almost $100$\% at a wavelength of 
$355$~$\mathrm{nm}$. One of the two mirrors reflects laser light in the vertical 
direction, and the other in the horizontal direction at a proper height to make 
collisions with the electron beam. Each mirror is mounted on a dual-axis automatic 
precision stage that rotates in two orthogonal directions. A combination of the two 
mirrors enables the direction change and parallel shift of an incident laser beam. 
The automatic precision stage can be remotely controlled from the outside of the 
laser injection room, and the mirror angle can be adjusted in increments of 
approximately $30$~$\mathrm{\mu rad}$.

A laser beam whose axis has been adjusted by the above two mirrors are further 
reflected in a direction parallel to the beamline at a reflective surface of 
a prism mirror. This mirror is made of a synthetic quartz right-angle prism with 
two orthogonal reflective surfaces, each of which has a size of 
$80$~$\mathrm{mm}$ $\times$ $80$~$\mathrm{mm}$. The reflective surface is coated 
exclusively for a wavelength of $355$~$\mathrm{nm}$. Two different laser beams at 
the same height are individually redirected by the two reflective surfaces, and 
merged into adjacent beams traveling in the same direction. There are two prism 
mirrors with slightly different heights to handle four laser injection by making 
two sets of side-by-side beams vertically adjacent. The four laser beams are 
simultaneously guided into the storage ring tunnel by using three large mirrors 
with a diameter of $150$~$\mathrm{mm}$. These large mirrors are made by applying 
a dual-wavelength coating for $355$ and $266$~$\mathrm{nm}$ on a circular substrate 
of synthetic quartz. The two prism mirrors and the three large mirrors are 
individually mounted on manual precision stages. The setting of their positions, 
heights, and angles was fixed after the adjustment in the initial stage of beamline 
operation. 

Inside the laser injection room, two of the three large mirrors and the merged 
laser-beam paths between them are shielded by a lead-plate box with a thickness of 
$3$~$\mathrm{mm}$. Thanks to this radiation shield, we can perform manual adjustment 
and maintenance works for lasers and optical devices in the laser injection room even 
during the photon beam production. The other large mirror is set in an interlocked 
radiation protection hatch, which is located outside the laser injection room and 
covers a $160$-$\mathrm{mm}$ diameter hole for laser beams entering the storage ring 
tunnel. This hatch is constructed by concrete walls in a small area of 
$1.5$~$\mathrm{m}$ $\times$ $1.7$~$\mathrm{m}$. For the hole drilled through a side 
wall of the storage ring tunnel, we put a motor-driven laser shutter made of iron 
with a thickness of $400$~$\mathrm{mm}$. It can be remotely closed inside the tunnel 
for radiation shielding when laser beams are not injected.

\begin{figure*}[t]
  \centering
  \includegraphics[width=12cm,bb=0 0 682 331]{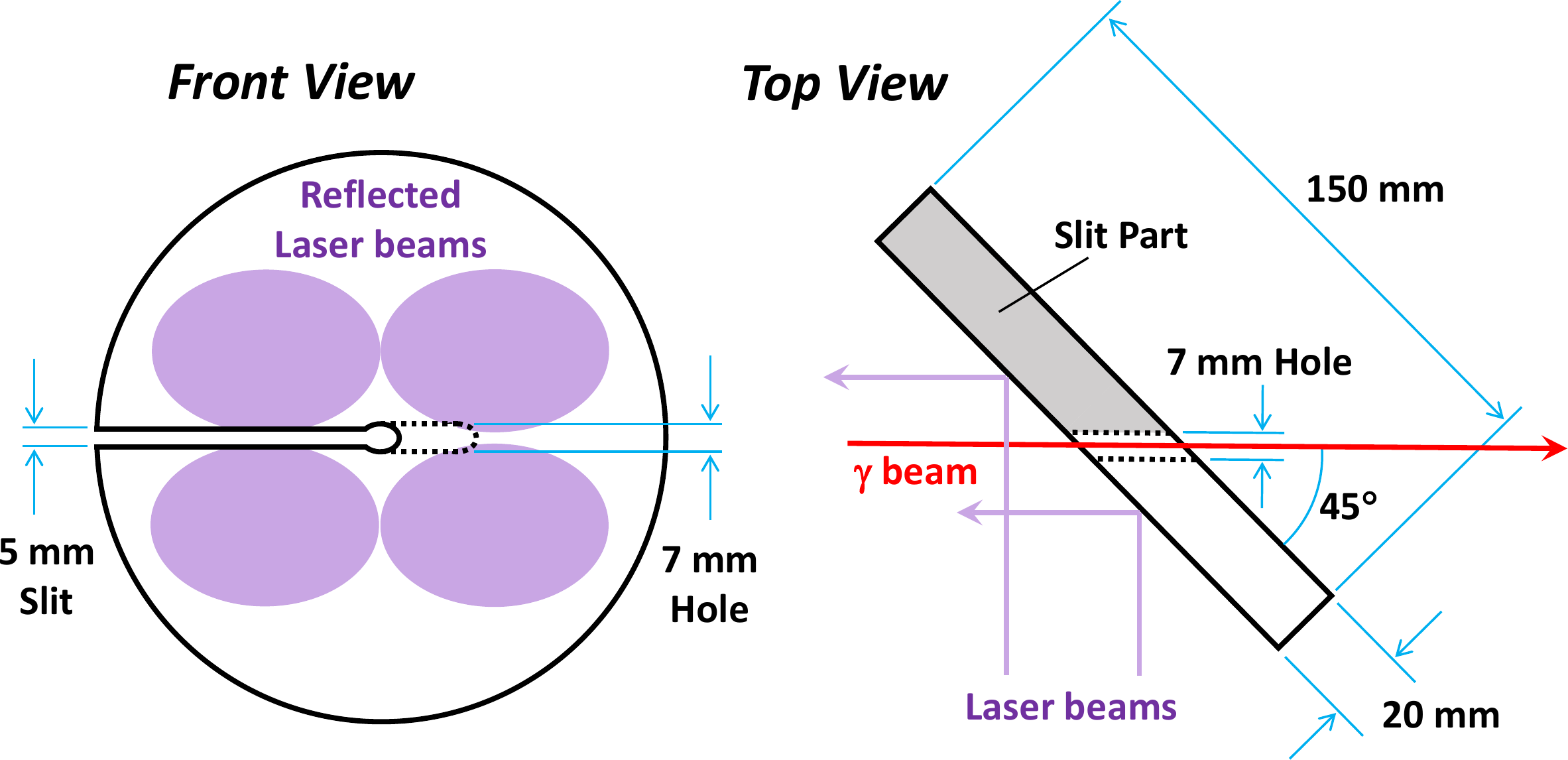}
  \caption{Front and top views of the first mirror installed in a front-end
           vacuum chamber. Four laser cross sections reflected at this mirror 
           are also indicated by smaller purple circles.}
  \label{fig:first_mirror}
\end{figure*}
In the storage ring tunnel, the incident laser light enters a front-end chamber 
from a side viewport with a ConFlat flange size of ICF203 (DN160). This viewport 
has a synthetic quartz window with an effective diameter of $140$~$\mathrm{mm}$ and 
a thickness of $8$~$\mathrm{mm}$. Its transmittance for UV wavelength light is about 
$94$\%. The incident laser light is then reflected toward the long straight section 
by the first mirror tilted $45$ degrees with respect to the beamline direction. 
The first mirror is made of an aluminum-coated silicon substrate with a diameter of 
$150$~$\mathrm{mm}$ and a thickness of $20$~$\mathrm{mm}$. Its reflectance is about 
$90$\% in a UV wavelength region. As shown in Fig.~\ref{fig:first_mirror}, 
a $5$-$\mathrm{mm}$ wide slit is horizontally machined at the electron beam height 
in the half area closer to the storage ring. This structure allows synchrotron 
radiation of $X$ rays to escape without heat load to the first mirror, whose 
reflective surface can be distorted by non-uniform temperature rise. Moreover, 
a water-cooled oxygen-free copper plate is applied to the back surface of the mirror 
in order to remove heat that is generated by a halo component of the synchrotron 
radiation hitting the edges of the slit. An entire mirror holder with the cooling 
plate is fixed by three columns that can be manually moved back and forth. The 
position and angle of the first mirror have been adjusted at the beginning of LEPS2 
beamline operation by using these columns.

At the first mirror, four incident laser beams are reflected in a square 
positional relationship as shown in Fig.~\ref{fig:first_mirror}. The reflected laser 
beams are focused at the Compton scattering point $22$~$\mathrm{m}$ ahead from the 
first mirror, and collide with the stored electron beam. A high-energy photon beam 
generated by Compton scattering travels back to the first mirror, and passes through 
its central hole with a diameter of $7$~$\mathrm{mm}$. This hole is made along the 
photon beam path from the center of the reflective surface with a tilting angle of 
$45$ degrees. This structure is important to eliminate the photon beam loss due to 
electron-positron pair creation.

The laser light is almost $100$\% linearly polarized when it is emitted from 
an oscillator. The direction of linear polarization is controlled by using an 
optical element called a $\lambda/2$ waveplate, which causes a phase difference 
of half a wavelength between two orthogonal polarization components. Because 
the LEPS2 facility uses high-power lasers, we have adopted a zero-order quartz 
waveplate with an air gap. In the laser injection room, a $\lambda/2$ waveplate 
is installed at the exit of each laser oscillator, and is rotated around the 
laser beam axis to adjust the direction of linear polarization horizontally or 
vertically. This rotational adjustment is done in the same way for all the laser 
beams that are simultaneously injected. The polarization direction is alternately 
changed about every week during the collection of physics data. The laser light 
immediately after the emission from an oscillator has a high power density, 
so that an irradiation position on the $\lambda/2$ waveplate is periodically 
shifted to avoid significant damage. For reducing such damage, a large 
$\lambda/2$ waveplate with a diameter of $60$~$\mathrm{mm}$ has been also 
introduced to control the polarization after magnifying a laser beam diameter 
by a beam expander. But the production cost of large waveplates is high and 
the installation of them is limited for a part of multiple laser beams. In the 
case of making circular polarization, an optical element called a $\lambda/4$
waveplate will be placed downstream of the $\lambda/2$ waveplate.

The polarization of a photon beam produced by Compton scattering is calculated 
by inputting that of laser light into Eqs.~(16) and (17) of Ref.~\cite{dangelo}
for linear and circular polarization states, respectively. The linear polarization 
is highest at the Compton edge, and gets lower as the photon beam energy decreases. 
The polarization of laser light acts as a scale factor for that of a photon beam 
in the above calculation. Here it is important to measure the linear polarization 
of laser beams just before Compton scattering because it may be deteriorated by 
reflections at many mirrors. This deterioration can happen by slightly different 
reflectances for the S- and P-wave components of laser light, which are defined 
by the direction of linear polarization with respect to the plane of incidence 
\cite{waveplate}. Therefore, the polarization measurement is regularly performed 
by inserting the monitor mirror on a laser beam path, as mentioned in 
Sec.~\ref{sec:design}, and extracting the laser light from a synthetic quartz 
viewport of a front-end vacuum chamber in the direction to the outer periphery 
of the storage ring.

The extracted laser light is incident on the polarization measurement system 
contained by a lead radiation shielding box in the storage ring tunnel. A main 
unit of the polarization measurement system consists of a polarizer (Glan Laser 
prism made of calcite) that rotates by remote control and a silicon photodiode 
(S8746-01 manufactured by Hamamatsu Photonics K.K.) that measures an intensity 
of the laser light after passing through the polarizer. A neutral-density filter 
for dimming is inserted between the polarizer and the photodiode. For linearly 
polarized laser light, the above unit measures a sine curve of the transmitted 
light intensity, where the maximum and minimum values ($P_\mathrm{max}$ and 
$P_\mathrm{min}$, respectively) appear alternately depending on the rotation 
angle of the polarizer. The linear polarization is given by the calculation of 
$(P_\mathrm{max} - P_\mathrm{min}) / (P_\mathrm{max} + P_\mathrm{min})$. In the 
operation so far, the measured polarization was typically about $98$\% and there 
was no siginificant difference between the horizontal and vertical polarization 
states. Because four laser beams that are simultaneously injected reach different 
positions in the polarization measurement system, the unit of the polarizer and 
the photodiode is mounted on a stage that moves to the position of each laser 
beam by remote control.

\section{Beamline chambers and transport line} \label{sec:bmline}

\begin{figure*}[t]
  \centering
  \includegraphics[width=16cm,bb=0 0 466 110]{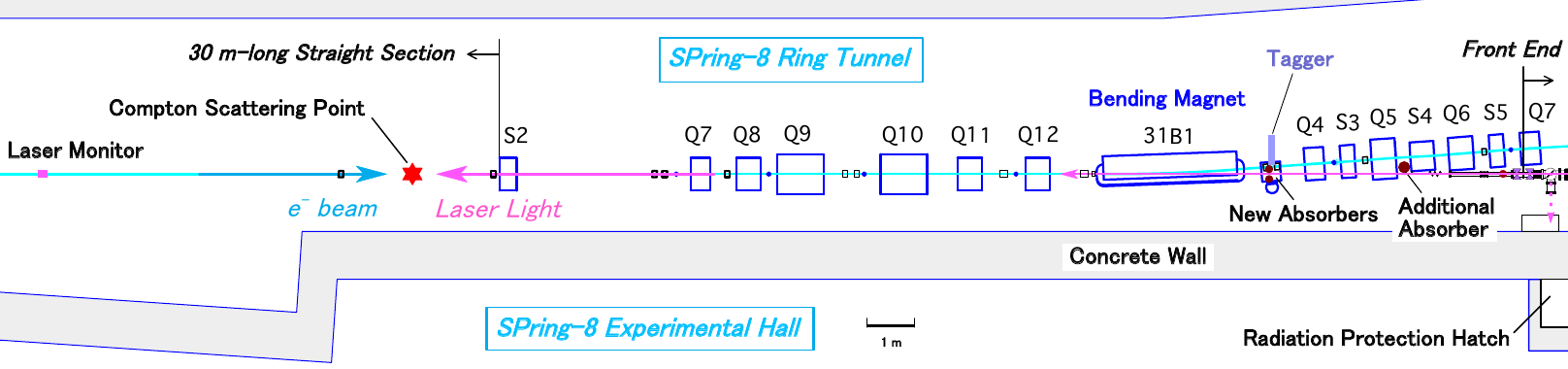}
  \caption{The upstream part of the LEPS2 beamline from the laser monitor to 
           the bending magnet, where the accumulated electron orbit and the 
           produced photon beam path are separated.}
  \label{fig:straight_section}
\end{figure*}
Figure~\ref{fig:straight_section} shows a plan view of the LEPS2 beamline from the 
downstream part of the long straight section to the vicinity of the tagger, which 
is located downstream of the bending magnet. The explanation of newly produced and 
modified chambers for the construction of the LEPS2 beamline can also be found in 
Ref.~\cite{yorita}. Laser Compton scattering is designed to occur at a place that 
the laser light reaches after entering the $30$-$\mathrm{m}$ long straight section 
and traveling a distance of $2$~$\mathrm{m}$. Most of the unscattered laser light 
goes upstream of the storage ring, but cannot escape from the viewport on the 
opposite side of the straight section, being repeatedly reflected by the inner 
walls of vacuum chambers. Such laser reflections inside the storage ring can cause 
vacuum deterioration that is not negligible in maintaining an ultra-high vacuum. 
Therefore, beamline baking with laser injection is regularly performed during the 
shutdown periods of the storage ring.

\begin{figure*}[t]
  \centering
  \includegraphics[width=14cm,bb=0 0 772 242]{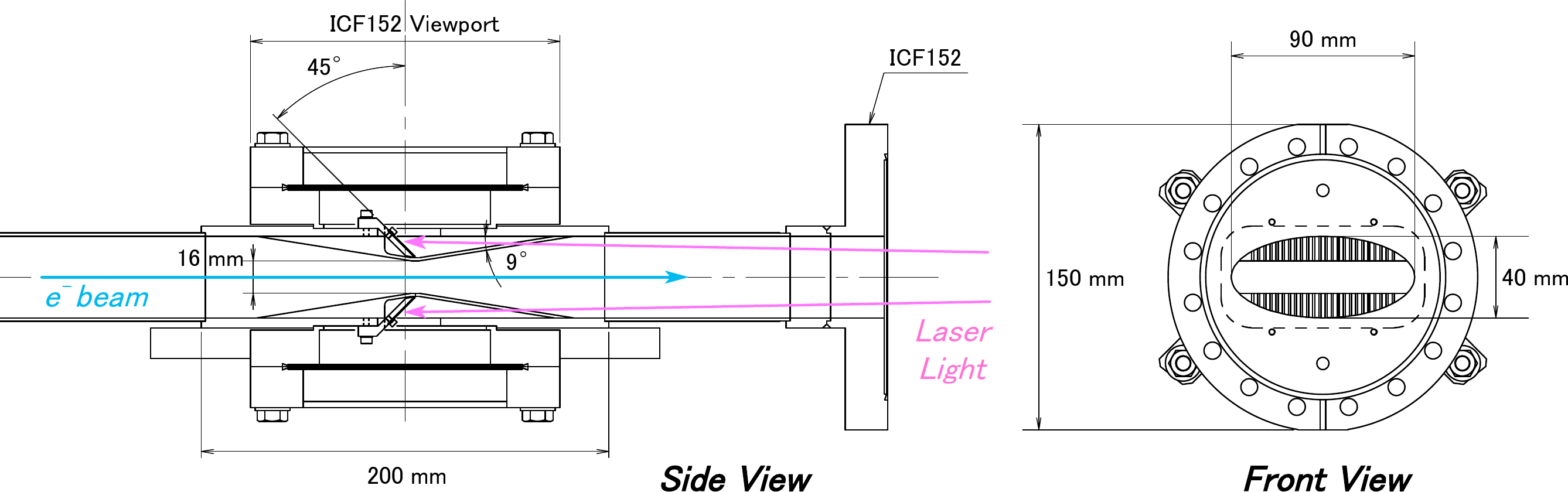}
  \caption{The laser monitor chamber installed $8.4$~$\mathrm{m}$ upstream of
           the Compton scattering point in the straight section of the LEPS2
           beamline. The ConFlat flange ICF152 (DN100) has an outer diameter 
           of $152$~$\mathrm{mm}$.}
  \label{fig:laser_monitor}
\end{figure*}

In order to know the directions of incident laser beams, it is necessary to install 
a dedicated chamber where laser beam spots can be observed before the inner reflections 
happen in the long straight section. This observation system is called the laser 
monitor. The $725$-$\mathrm{mm}$ long chamber that was originally located 
$8.4$~$\mathrm{m}$ upstream of the Compton scattering point is replaced with 
the laser monitor chamber shown in Fig.~\ref{fig:laser_monitor}. Inside this chamber, 
comb-shaped windows with many $1$-$\mathrm{mm}$ wide slits are attached at an angle 
of $9$ degrees on the upper and lower sides not to interfere with the surface current 
that accompanies the stored electron beam. A distance to the electron beam orbit is 
$8$~$\mathrm{mm}$ at the closest points of the window structures. The laser light 
that has passed through either of the upper or lower slits hits an internal screen 
made of an alumina fluorescent plate, allowing the irradiated position to be observed 
from a CCD camera set in the atmosphere outside a viewport. 

A high-energy photon beam produced by laser Compton scattering goes back on the path of 
the incident laser light in the opposite direction. After leaving the $30$-$\mathrm{m}$
long straight section, the photon beam passes through straight chambers with quadrupole, 
sextupole, steering magnets, etc.~for a distance of $10$~$\mathrm{m}$. Then, the photon
beam path is separated from the accumulated electron orbit in a vacuum chamber set 
inside a gap of the bending magnet. In this chamber, the aperture for the photon beam
path is widened as much as possible, as described in Sec.~\ref{sec:design}.

Highly brilliant synchrotron radiation of $X$ rays is not only an unnecessary 
contamination for the $\mathrm{GeV}$ $\gamma$-ray beam but also a source of heat 
generation and vacuum deterioration due to their scattering at vacuum chambers and 
optical devices. Thus, two new $X$-ray absorbers are installed in the chamber just 
downstream of the bending magnet. They are attached instead of a conventional crotch 
absorber, and cover side areas of the photon beam path so as not to interfere with 
the laser injection; One is placed on the accumulated electron orbit side to receive 
brilliant synchrotron radiation, while the other is put on the outer circumference 
side to absorb a halo component of $X$ rays. For the chamber to which the new 
absorbers are mounted, further modification has been made on the inner side of the 
accumulated electron orbit; The chamber size is horizontally enlarged to secure the 
widespread path of recoil electrons, which lose energy by Compton scattering and are 
largely bent at the bending magnet. A horizontal-slit window made of $3$-$\mathrm{mm}$ 
thick aluminum-alloy is attached to the modified chamber as the exit of recoil 
electrons. Recoil electrons extracted into the atmosphere are analyzed by the tagger 
to measure their momenta, as explained in Sec.~\ref{sec:tagger}.

After passing through the bending magnet, the photon beam further travels inside
large-aperture straight chambers located in a $4$-$\mathrm{m}$ long area beside 
a series of convergent magnets. Here, an additional $X$-ray absorber is installed 
to stop the horizontally spreading synchrotron radiation. This absorber is shaped 
in a narrow fin extended from the accumulated electron orbit side at the photon 
beam height. It is designed to avoid obstructing the laser injection but block
a part of synchrotron radiation.

The photon beam then enters the section of front-end chambers which have the optical 
devices related to the laser injection. At an upstream chamber of the front-end section, 
the monitor mirror made of a $20$-$\mathrm{mm}$ thick silicon substrate can be inserted 
downward, as mentioned in the previous sections. The mirror substrate is horizontally 
tilted by $45$ degrees to the beamline. It also serves as an absorber to prevent 
synchrotron radiation from penetrating downstream of the beamline when a photon beam 
is not generated. This mirror absorber can be remotely raised or lowered by compressed 
air operation when necessary. 

The front-end vacuum section ends $2$~$\mathrm{m}$ downstream of the first mirror or 
$24.1$~$\mathrm{m}$ downstream of the Compton scattering point. The photon beam is once 
taken out into the atmosphere from a water-cooled ICF152 (DN100) flange with an aluminum 
window that is made thin down to $2$~$\mathrm{mm}$ in a $10$-$\mathrm{mm}$ diameter 
region. On the vacuum side of this window, an oxygen-free copper mask is attached with 
a water-cooling pipe. This mask receives synchrotron radiation in a rectangular part 
that has vertical and horizontal widths of $10$ and $120$~$\mathrm{mm}$, respectively, 
with depth-direction slopes toward a narrow central area. The central area corresponds 
to a photon beam path with the thin aluminum window. A vacuum in the front-end section 
is maintained in the range around $10^{-8}$--$10^{-7}$~$\mathrm{Pa}$ by two ion and 
two titanium getter pumps.

The photon beam extracted into the atmosphere immediately passes through an $X$-ray 
absorber made of a $1$-$\mathrm{mm}$ thick tungsten plate and a $7$-$\mathrm{mm}$ 
diameter lead collimator with a length of $100$~$\mathrm{mm}$. In the LEPS2 beamline, 
the straight section common to the storage ring is much longer than that of the usual 
beamlines, so that there is a non-negligible amount of the contamination due to 
bremsstrahlung $\gamma$ rays generated by accumulated electrons passing through 
a residual gas inside the ring. Therefore, a broad angular component of $\gamma$ rays 
are cut off by the collimator. Lower-energy $\gamma$ rays have angular broadening in 
the kinematics of laser Compton scattering, so a component of the photon beam below 
about $0.5$~$\mathrm{GeV}$ cannot go through the lead collimator, as recognized from 
Fig.~\ref{fig:beam_divergence}. 

A sweep magnet, which is a $1$-$\mathrm{m}$ long neodymium magnet with a magnetic 
field of $0.63$~$\mathrm{T}$, is installed downstream of the lead collimator to remove 
the contamination of electrons and positrons produced by pair creation at the upstream 
materials. At a location $10$~$\mathrm{m}$ downstream of the lead collimator, an iron 
collimator with a diameter of $11$~$\mathrm{mm}$ and a length of $200$~$\mathrm{mm}$ 
is placed to serve as a buffer collimator to shield secondary particles produced with 
large angles at the tungsten absorber and the aperture edge of the lead collimator. 
Downstream of the iron collimator, we put a beam shutter to provide radiation shielding 
when a photon beam is not generated. All the devices described above are located inside 
the $1$-$\mathrm{m}$ thick concrete walls of the storage ring tunnel so as to avoid 
possible problems on radiation protection.

A photon-beam transport pipe with a medium vacuum of $10$~$\mathrm{Pa}$ starts 
immediately downstream of the sweep magnet, and runs $95.6$~$\mathrm{m}$ through 
the concrete wall of the ring tunnel and the experimental hall in the storage ring 
building. It extends to the LEPS2 experimental building, which is independently 
constructed for hadron photoproduction experiments. The inner diameter of the 
transport pipe is $60$~$\mathrm{mm}$ in the storage ring tunnel and gradually 
increases to $200$~$\mathrm{mm}$ as it approaches the LEPS2 experimental building. 
On both ends of the transport pipe, thin films are attached as a partition between 
the atmosphere and a vacuum. A Kapton film with a thickness of $75$~$\mathrm{\mu m}$ 
($125$~$\mathrm{\mu m}$) and an aramid film with a thickness of $50$~$\mathrm{\mu m}$ 
($50$~$\mathrm{\mu m}$) are superimposed for the upstream (downstream) end. For 
putting the films safely on the downstream end of the transport pipe, its inner 
diameter is converted from $200$~$\mathrm{mm}$ to $100$~$\mathrm{mm}$ by adding 
a short straight pipe.

\section{Experimental building} \label{sec:expbldg}

\begin{figure*}[t]
  \centering
  \includegraphics[width=14cm,bb=0 0 717 481]{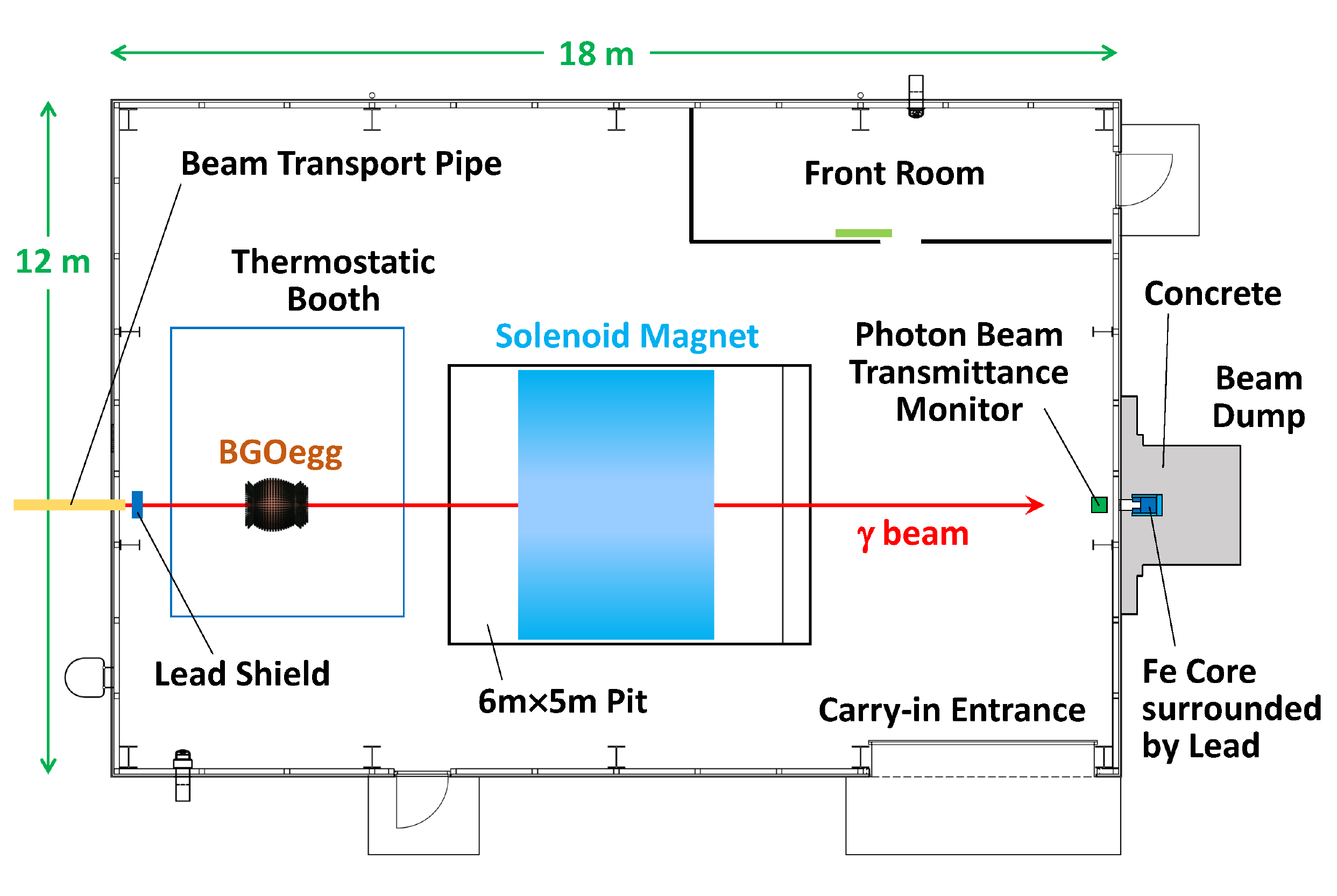}
  \caption{A plan view of the LEPS2 experimental building, located in the downstream 
           end of the LEPS2 beamline.}
  \label{fig:leps2_building}
\end{figure*}

The LEPS2 experimental building is located at the most downstream of the LEPS2 
beamline and has enough space to contain large detector systems covering most of 
all solid angles. The size of the LEPS2 experimental building is $18$~$\mathrm{m}$ 
in the direction of a photon beam, $12$~$\mathrm{m}$ in width, and $10$~$\mathrm{m}$ 
in height. It is equipped with an overhead traveling crane allowing a maximum load 
of $2.8$~$\mathrm{t}$. Figure~\ref{fig:leps2_building} shows a floor plan of the 
LEPS2 experimental building. In the future, this building can be expanded downstream 
to an area about $1.5$ times larger. The downstream end of the photon-beam transport 
pipe is located at the most upstream part of the LEPS2 experimental building. 
A $200$-$\mathrm{mm}$ thick lead shield with a $45$-$\mathrm{mm}$ diameter hole 
is installed just downstream of the pipe end. This hole is slightly wider than the 
maximum spread expected for the photon beam that is cut off by the $7$-$\mathrm{mm}$ 
diameter collimator in the storage ring tunnel. It prevents a halo component of the 
photon beam from entering the downstream detector systems. Further downstream but 
just upstream of the individual detector systems, additional lead shields or 
collimators are placed depending on the experiments. Charged particles contaminating 
the photon beam or generated by the halo component of the beam are rejected in the 
offline analysis by using a large plastic scintillator, called the UpVeto counter.

A beam dump, which receives all photons, is located at the downstream end of the 
LEPS2 experimental building \cite{bmdump}. A core of the beam dump is made of iron 
with a cross section of $400$~$\mathrm{mm}$ $\times$ $400$~$\mathrm{mm}$ and a depth 
of $300$~$\mathrm{mm}$. In front of the core, a $150$-$\mathrm{mm}$ thick iron block 
having a $100$-$\mathrm{mm}$ square hole along the photon beam axis is attached to 
absorb the backscatter of electromagnetic showers. The back and sides of those iron 
blocks are covered with $100$-$\mathrm{mm}$ and $50$-$\mathrm{mm}$ thick lead, 
respectively. The surrounding volume with a horizontal width of $2200$~$\mathrm{mm}$, 
a height of $2600$~$\mathrm{mm}$, and a beam-direction depth of $1800$~$\mathrm{mm}$ 
is further filled with concrete. This structure allows a photon-beam intensity 
(a Compton scattering rate over all the energy range) up to 
$5 \times 10^7$~$\mathrm{s^{-1}}$ in terms of the radiation safety. In addition, 
the outside area of the front part of the beam dump is reinforced with 
a $300$--$400$-$\mathrm{mm}$ thick concrete shield wall to stop particles photoproduced 
with angular spread from the fixed target inside a detector system. The additional 
concrete wall is designed to permit a target thickness to be increased up to $0.1$ 
radiation length in the LEPS2 experimental building. Almost all of the LEPS2 
experimental building is set to a radiation controlled area, except for the front 
room with a size of $2.4$~$\mathrm{m}$ $\times$ $7.5$~$\mathrm{m}$ near the entrance. 
The amount of radiation leakage is reduced by taking a large space whose boundary is 
far enough from the fixed target and restricting access to this area with an interlock 
system.

Different detector systems are separately set up in the upstream and middle 
parts of the LEPS2 experimental building, and operated in different periods
of the beamtime depending on experimental programs. Details of the individual 
detector systems will be described in separate articles. In the upstream area, a 
large-acceptance electromagnetic calorimeter ``BGOegg'', which covers polar angles 
from $24$ to $144$ degrees, is installed with a fixed target at a distance of 
$125$~$\mathrm{m}$ from the Compton scattering point. This calorimeter consists 
of 1,320 BGO crystals assembled in an egg shape, and a total length of the detector 
reaches about $1$~$\mathrm{m}$ between the upstream and downstream ends. It excels 
in detecting photoproduced neutral mesons, decaying into multiple $\gamma$ rays. The 
energy resolution for $1$-$\mathrm{GeV}$ $\gamma$ rays is $1.3$\% \cite{eggdesign}, 
corresponding to the highest performance in the world. The BGOegg is combined with 
charged-particle detectors that are installed onto its inner and forward sides to 
detect and identify all the final states of a hadron photoproduction reaction as 
much as possible. Because of temperature dependence in the amount of scintillation 
light of a BGO crystal and the signal output of a photomultiplier tube, the entire 
detector system is housed in a thermostatic booth with an area of 
$5.2$~$\mathrm{m}$ $\times$ $4.2$~$\mathrm{m}$. The detector system with the BGOegg 
calorimeter is already under operation. Physics data-taking using a liquid hydrogen 
or carbon target was done from 2014 to 2016. So far, we have shown results in the 
analyses searching for baryon resonances and mesic nuclei \cite{eggpi0, etapnuc, eggomg}.

In the center of the LEPS2 experimental building, there is a pit with an area 
of $5$~$\mathrm{m}$ $\times$ $6$~$\mathrm{m}$ and a depth of $1.5$~$\mathrm{m}$. 
The center of the pit is $131$~$\mathrm{m}$ away from the Compton scattering 
point. A large solenoid magnet weighing $400$~$\mathrm{t}$ with a diameter of 
$5$~$\mathrm{m}$ and a length of $3.5$~$\mathrm{m}$ has been relocated to this pit 
from the E787/E949 experiment \cite{e787e949} at Brookhaven National Laboratory.
The central axis of the solenoid magnet is aligned along the photon beam. 
Inside the $2.96$-$\mathrm{m}$ diameter bore with a central magnetic field of 
$1$~$\mathrm{T}$, we have set up a large charged-particle spectrometer, which 
consists of a time projection chamber, four planar drift chambers, and many types 
of particle identification detectors \cite{solexp}. Data acquisition using a liquid 
hydrogen or deuterium target is beginning for studying the photoproduction of 
a pentaquark $\Theta^+$, which is attracting attention as an exotic hadron 
\cite{theta1,theta2}, and a hyperon $\Lambda (1405)$, which is expected to have 
a hadronic molecular structure \cite{l1405a, l1405b}. The use of a large solenoid 
magnet with normal conductivity scales up the infrastructure of the LEPS2 experimental 
building by requiring a unit receiving and transforming an electric power up to 
$2$~$\mathrm{MVA}$ and a cooling water system with a capacity of $1.5$~$\mathrm{MW}$ 
as ancillary equipment.

\section{Photon tagger} \label{sec:tagger}

The tagger detects recoil electrons whose energy is lost by laser Compton scattering.
As shown in Fig.~\ref{fig:straight_section}, it is installed near the exit of the 
$0.68$-$\mathrm{T}$ bending magnet downstream of the long straight section. It is 
located on the inner circumference side of the 
storage ring to measure the momentum of a recoil electron. From the four-momentum 
conservation in laser Compton scattering, the measured electron momentum is converted 
to the energy of a backscattered photon event by event. In addition, the integrated 
number of scattered photons during an experiment is obtained by counting the number 
of recoil electrons detected at the tagger, and used to measure the cross sections 
of hadron photoproduction reactions. The detection of a recoil electron also serves 
to generate a trigger signal for physics data acquisition.

\begin{figure*}[t]
  \centering
  \includegraphics[width=13cm,bb=0 0 294 178]{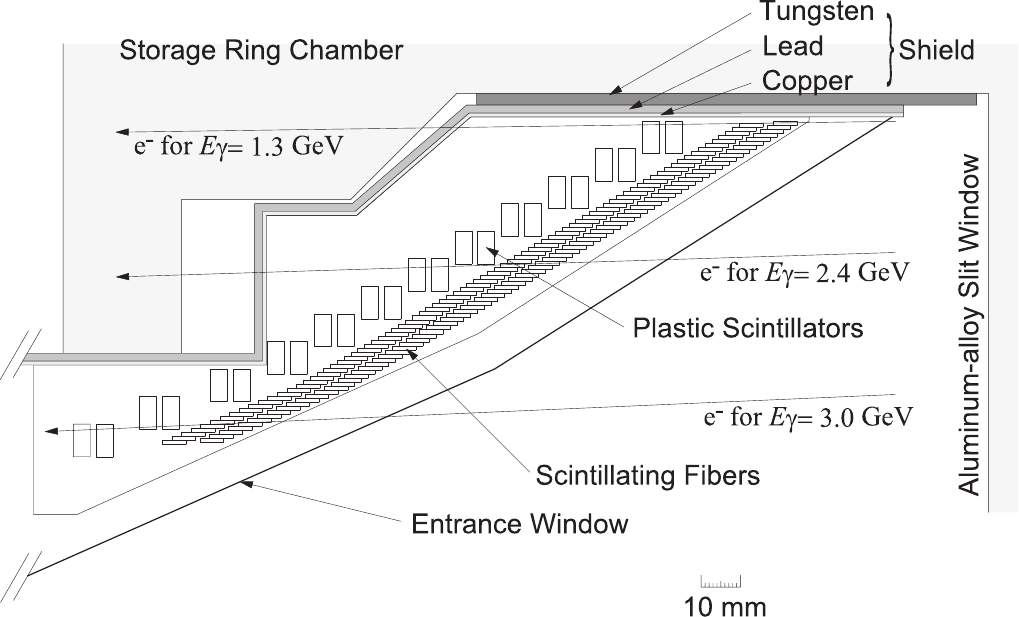}
  \caption{A plan view of the tagger with two layers of scintillating fibers
           (upstream) and two layers of plastic scintillators (downstream).}
  \label{fig:tagger_structure}
\end{figure*}
The detector structure of the tagger is shown in Fig.~\ref{fig:tagger_structure}. 
Two layers of bundles made of $1$-$\mathrm{mm}$ square scintillating fibers are 
arranged in the upstream side, while two layers of plastic scintillators having 
a width of $8$~$\mathrm{mm}$, a thickness of $4$~$\mathrm{mm}$, and a height of 
$10$~$\mathrm{mm}$ are placed in the downstream side. Six scintillating fibers 
are bundled into a single readout channel by lining them up in the direction of
a recoil electron track to increase a detection efficiency. The recoil electron 
whose trajectory has been bent at the bending magnet changes a hit position in the 
tagger depending on its momentum. The digitization of hit positions is subdivided
by installing the two fiber-layers with a $0.5$~$\mathrm{mm}$ shift to each other. 
Correspondingly, the measurement of a photon beam energy is digitized by about 
every $15$~$\mathrm{MeV}$, resulting in the detector resolution of 
$15/\sqrt{12}=4.3$~$\mathrm{MeV}$. 

The plastic scintillators in the downstream side are arranged so that two channels 
in different layers should be paired with geometrical overlap in the direction of 
a recoil electron path. A pair of plastic scintillators covers an about 
$240$~$\mathrm{MeV}$ range of the photon beam energy. The $2$-$\mathrm{mm}$ wide 
region at the edge of each scintillator overlaps with that of an adjacent scintillator. 
The overlapped region corresponds to a photon beam energy range of about 
$60$~$\mathrm{MeV}$. If a recoil electron passes through this region, four plastic 
scintillators have hit signals.

The position of individual scintillating fiber bundles and plastic scintillators in 
the same layer is shifted downstream as they are away from the electron 
beam orbit toward the inner side. This arrangement is adopted because a large amount 
of $X$ rays by synchrotron radiation are scattered at an absorber in the vacuum chamber 
immediately downstream of the bending magnet. The entire tagger is enclosed in a shield 
box to prevent these $X$ rays from entering the signal-sensitive parts. The layers of 
scintillating fibers and plastic scintillators are arranged in a shadow of the shield-box 
wall on the storage ring side. The shield box is made of $1$-$\mathrm{mm}$ thick copper 
and $2$-$\mathrm{mm}$ thick lead plates. A part of the shielding wall where $X$ rays are 
particularly intense on the storage ring side is made of a $3$-$\mathrm{mm}$ thick 
tungsten plate instead of the lead plate. Only in the area that recoil electrons go
through, a narrow window is left on the upstream wall of the shield box.

A 16-channel multi-anode photomultiplier tube (H6568-200MOD by Hamamatsu Photonics 
K.K.) and a metal-package photomultiplier tube (R9880U-210 by the same manufacturer) 
are used to read signals from scintillating fiber bundles and a plastic scintillator, 
respectively. For the readout of a scintillating fiber signal, only the detection time 
of a recoil electron is recorded in data collection, and the gain of a photomultiplier 
tube is adjusted before the data-taking by determining the applied voltage to an optimum 
value giving an enough detection efficiency. On the other hand, both the detection time 
and energy are recorded as data for a plastic scintillator signal. 

A recoil electron track is reconstructed under the conditions that one or two layers 
of scintillating fibers and both layers of plastic scintillators have coincident hits 
and that the positions of hit fibers and scintillators are geometrically on a straight 
line nearly parallel to the accumulated electron orbit. The hit timing of a plastic 
scintillator is tightly required to be consistent with that of a paired plastic 
scintillator in the other layer. The detection time of a recoil electron track is 
calculated by averaging the hit timings of plastic scintillators. When multiple 
recoil-electron tracks are found in one recorded event, only one track that is 
associated with a physics reaction can be selected by requiring its detection time 
to be consistent with the reaction timing determined by a detector system in the 
LEPS2 experimental building. However, the track selection is impossible if multiple 
recoil-electron tracks are detected without a time difference larger than 
$2$~$\mathrm{ns}$, which corresponds to the minimum bunch interval of the stored 
electron beam. In this case, a true photon beam energy cannot be identified for 
a hadron photoproduction event. Such an event was discarded in physics analyses.

Recoil electrons that have not lost large energy at Compton scattering pass near the 
accumulated electron orbit without entering the tagger. Therefore, there is a lower 
limit to the photon beam energy that can be measured. It is $1.26$~$\mathrm{GeV}$ for 
the case of the LEPS2 beamline. Calibration of the photon-beam energy determination
at the tagger was performed by comparing the hit position of a recoil electron with 
a photon energy predicted in an independent measurement. Here, the detector system 
in the LEPS2 experimental building can measure all the final-state particles of 
a hadron photoproduction event, and it is possible to evaluate a photon beam energy 
from the kinematic fit that does not use the tagger information. In this energy 
calibration, a quartic function was fitted to the correlation between the predicted 
energy and the hit fiber position. This fit was constrained by an additional condition 
that the polynomial function should represent the maximum photon beam energy, calculated 
based on Eq.~\ref{eqn:maxene}, at the Compton edge observed in the distribution of hit 
fiber positions.

The reconstruction efficiency for a recoil electron track is mainly determined by 
the following three sources: $\rm(\hspace{.18em}i\hspace{.18em})$ a percentage of 
a straight track that is successfully reconstructed from a combination of 
scintillating fibers and plastic scintillators, $\rm(\hspace{.08em}ii\hspace{.08em})$ 
a rate to find only one track that can be separated at a reaction time for a certain 
event, and $\rm(i\hspace{-.08em}i\hspace{-.08em}i)$ detection efficiencies of the 
scintillating fibers for which discriminator thresholds become effectively high 
when the gains of photomultiplier tubes drop. This efficiency is evaluated by using 
a data sample in which the photon beam energy can be obtained independently of the 
tagger information, as done in the energy calibration. The efficiency value varies 
depending on the hit position of a tagger track or the photon beam energy because of 
the source $\rm(i\hspace{-.08em}i\hspace{-.08em}i)$. Its typical value is estimated 
to be around $90$\%.

A tagger trigger signal, which is a logic signal with a width of $20$~$\mathrm{ns}$, 
is generated when there is at least one pair of coincident hits at geometrically 
overlapped channels in the two layers of plastic scintillators. This signal is 
incorporated as a part of the hadron photoproduction trigger required for physics 
data acquisition. It is also used to measure the intensity of a tagged photon beam 
by counting the number of signals at a scaler. The integrated number of tagger 
trigger signals is important for calculating the luminosity of collected data. In 
the measurement of a photon beam intensity, it is not possible to separate time 
difference equal to or less than the width of the tagger trigger signal, so it is 
necessary to correct dead time. This correction is purely stochastic, and influenced 
by the filling pattern of electron beam bunches at SPring-8 \cite{filling}. The amount 
of dead time correction increases as a length of bunch trains, where electron bunches 
are filled every $2$~$\mathrm{ns}$ according to the radio frequency applied to the 
accelerator, becomes longer. In contrast, the amount of correction is smaller when 
electron bunches are filled at equal intervals every several tens of $\mathrm{ns}$. 
Moreover, a larger dead time correction is needed as the photon beam intensity 
itself increases. Such correction factors are obtained by a Monte Carlo simulation 
depending on the filling pattern and the beam intensity. 

The tagger trigger counts are affected by the contamination originating from
high-momentum recoil electrons, which take the paths closer to the accumulated 
electron orbit and thus should not be tagged in principle. A part of these recoil 
electrons hit a side wall of the tagger shield box or a vacuum chamber structure 
to extract recoil electrons into the tagger. The electron hit can cause an 
electromagnetic shower that leaves signals at the sensitive parts of the tagger. 
A contamination rate of the shower contribution in the tagger trigger counts is 
estimated to be $4.2$\% for the LEPS2 beamline. This rate is much lower than 
that of the LEPS beamline ($\sim 21$\%) because we have cut off a wall structure 
separating the storage ring and the space expanded for recoil electron paths inside 
the vacuum chamber just upstream of the tagger. In the offline reconstruction of 
a recoil electron track, the contamination of an electromagnetic shower is removed 
by tightening the conditions of geometrical correspondence among hits in multiple 
layers and limiting the number of hit fibers that can be continuously connected 
as a cluster in a certain layer. 

\section{Photon beam properties} \label{sec:property}

The construction of the LEPS2 beamline started in 2010, and the first observation of 
photon beam production was successfully done in January 2013 \cite{baryons2013}. 
After commissioning, the first phase of full-scale data collection was carried 
out from 2014 in the experiments using the BGOegg calorimeter. In the operation 
so far, only UV lasers with a wavelength of $355$~$\mathrm{nm}$ have been used.
Depending on experimental periods, two to four laser beams were simultaneously 
injected into the storage ring. The number of lasers in use often decreases from 
the maximum of four because some of oscillators alternately need to be overhauled 
for about half a year due to the deterioration of inner optics by aging. We 
currently operate one and two high-power lasers with outputs of $24$~$\mathrm{W}$ 
and $16$~$\mathrm{W}$, respectively, and additionally use old oscillators with 
an output power of $8$~$\mathrm{W}$ at remaining slots for multiple laser injection. 
The optical system for laser injection is damaged due to the long-term irradiation 
of laser light and the leakage of $X$ rays by synchrotron radiation. Since the 
transmittance and reflectance of optical devices decrease, they should be renewed 
as appropriate. In particular, the first mirror and synthetic quartz window 
installed in a front-end vacuum chamber need to be replaced once every few years. 
For deep-UV lasers, we will proceed with their installation in the future after 
the development of high-power oscillators has been achieved, as described in 
Sec.~\ref{sec:summary}.

The intensity of a photon beam reaching a target for hadron photoproduction 
experiments in the LEPS2 experimental building is estimated by correcting the 
counting rate of tagger trigger signals with the simulated dead time and the 
shower contamination rate, explained in Sec.~\ref{sec:tagger}. Then, the 
corrected tagger rate must be multiplied by the beamline transmittance for 
photons traveling a distance of about $130$~$\mathrm{m}$. A time-integrated 
value of the photon beam intensity is important for obtaining the differential 
cross sections of hadron photoproduction reactions measured in the detector 
system around the target. As already mentioned, there are various materials 
on the photon beam path from the Compton scattering point to the fixed target 
in the LEPS2 experimental building, causing conversions to electron-positron 
pairs. Here, the beamline materials refer to the aluminum window at the exit 
of the front-end section, the tungsten $X$-ray absorber, the Kapton and aramid 
films attached to both ends of the photon-beam transport pipe with a medium 
vacuum, the air space after leaving the front-end section or the beam transport 
pipe, and the UpVeto counter. The transmittance calculated from the amount of 
these materials is $77.2$\%, where the influence of the $1$-$\mathrm{mm}$ thick 
tungsten plate is the most effective. On the other hand, the LEPS beamline has 
a structure in which a photon beam passes through a silicon mirror in a vacuum 
\cite{nimlepsbl}, providing a lower transmittance of about $67$\%. 

Recently, a system to measure the photon beam transmittance by counting a rate 
of pair creation with a converter has been developed to be installed upstream 
of the beam dump in the LEPS2 experimental building \cite{hasaito}. The ratio 
of counting rates in this system and the tagger will be immediately monitored 
during experimental periods. The measured transmittance is in agreement with 
the value calculated from the amount of materials. So far, we have observed 
a phenomenon in which the transmittance depends on the photon beam energy when 
the focal point of incident laser light shifts farther than the design, causing 
the cut-off of a lower-energy and more-widespread component of the photon beam 
at the lead collimator \cite{eggpi0, hasaito}. Therefore, it is important to 
monitor the laser focal point through the confirmation of no existence of 
energy dependence in the transmittance measured with this system.

In the LEPS2 beamline, the maximum rate of tagged photons, obtained based on the 
tagger trigger rate with the corrections for dead time and shower contamination, 
has reached $3.0 \times 10^6$~$\mathrm{s^{-1}}$ by injecting the UV laser beams 
into the storage ring with an electron beam current of $100$~$\mathrm{mA}$. After 
multiplied by the beamline transmittance, this rate corresponds to a photon beam 
intensity of $2.3 \times 10^6$~$\mathrm{s^{-1}}$ in the LEPS2 experimental 
building. The photon beam intensity at an experimental setup has increased by 
approximately $25$\% compared with the case that two UV laser beams are injected 
in the LEPS beamline. At the moment, the tagger trigger rate in the 
LEPS2 beamline is not as high as initially expected. This problem arises from the 
difficulty to precisely adjust the optimum axes of incident laser beams due to 
the deterioration of a vacuum during the storage ring operation. The vacuum 
deterioration occurs by repeated reflections of the laser beams inside the storage 
ring chambers, so that the beamline baking must be performed by laser injection with 
fixed beam axes before the ring operation, as explained in Sec.~\ref{sec:bmline}. 
We plan further improvement of the tagger trigger rate to be realized after 
the introduction of pulsed lasers, described in Sec.~\ref{sec:summary}. 

When laser beams are not injected, $\gamma$ rays by bremsstrahlung from the residual 
gas in the electron storage ring are generated in the form of a beam. In the LEPS2 
beamline, a portion that is common with the storage ring is as long as $50$~$\mathrm{m}$, 
including the $30$-$\mathrm{m}$ long straight section without any magnets and the 
additional $10$-$\mathrm{m}$ long parts at both ends with convergent magnets. So the 
tagger trigger rate by bremsstrahlung $\gamma$ rays becomes ($1$--$2$) $\times$
$10^4$~$\mathrm{s^{-1}}$, which is an order of magnitude more than that in the LEPS 
beamline with a straight section of $7.8$~$\mathrm{m}$. When laser beams are injected, 
the bremsstrahlung $\gamma$ rays are mixed in a tiny fraction with the photon beam 
generated by Compton scattering.

\begin{figure}[t]
  \centering
  \includegraphics[width=8cm,bb=0 0 491 463]{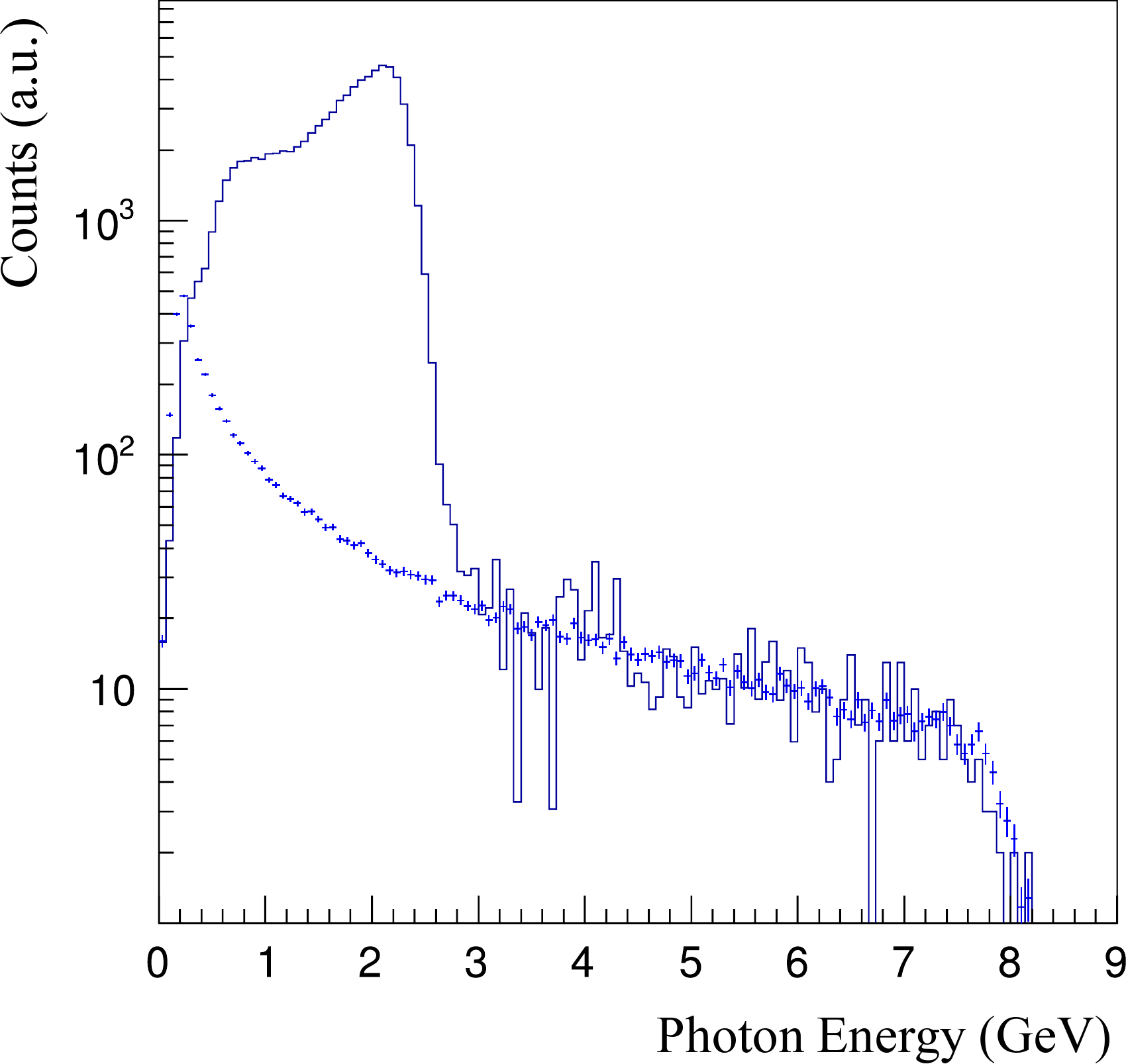}
  \caption{Photon-beam energy spectra measured by a lead-glass counter. 
           Two spectra when laser injection is turned on (dark-blue histogram) 
           and off (blue crosses) are overlaid.}
  \label{fig:energy_spectrum}
\end{figure}
The energy spectrum of a photon beam can be directly measured by placing a small 
detector on the photon beam axis in the LEPS2 experimental building. This measurement 
provides a smooth spectrum over all the energies in contrast to the indirect measurement 
by the tagger, which can only determine the energies higher than $1.26$~$\mathrm{GeV}$ 
and returns digitized values corresponding to the individual scintillating fiber 
channels. Figure~\ref{fig:energy_spectrum} shows photon-beam energy spectra obtained 
by using a SF5 lead-glass counter \cite{ins, forest} in the LEPS2 experimental building. 
The dimensions of the lead-glass counter are $150 \times 150$~$\mathrm{mm^2}$ in area 
and $300$~$\mathrm{mm}$ in length so as to catch an electromagnetic shower without 
leakage. Data was taken by self-triggers of the lead-glass counter. The photon beam 
intensity was intentionally lowered down to a tagger trigger rate of 
$0.7 \times 10^6$~$\mathrm{s^{-1}}$ for suppressing pile-up signals. The two spectra 
with and without laser injection were normalized by a gas-bremsstrahlung component 
in the energy range above $5$~$\mathrm{GeV}$. A small contribution from pile-up 
signals, appearing mainly in the energy range below $5$~$\mathrm{GeV}$, was subtracted 
by using a spectrum for events where multiple hits were clearly found in a short time 
range. In the case that laser light is injected into the storage ring, a component of 
$\gamma$ rays produced by Compton scattering is observed with a strength that is 
roughly two orders of magnitude higher than that by bremsstrahlung. It is also seen 
that the energy region lower than around $0.5$~$\mathrm{GeV}$ is cut off by the effect 
of the lead collimator with a diameter of $7$~$\mathrm{mm}$.

A photon beam energy measured for each event by using the tagger is fluctuated around
a true value with a finite resolution. By shifting the relative positions of the two 
layers of scintillating fibers, recoil electrons can be detected with an accuracy of 
each $0.5$~$\mathrm{mm}$, which corresponds to a tagger energy resolution of 
$4.3$~$\mathrm{MeV}$, as described in Sec.~\ref{sec:tagger}. However, the actual 
energy resolution has been evaluated to be $12.1$~$\mathrm{MeV}$ by comparing a photon 
beam energy measured by the tagger with an independent value predicted from the kinematic 
fit for a coincident photoproduction reaction. This resolution was confirmed by an 
alternative method where a complementary error function was fitted to the Compton edge 
of the energy spectrum measured by the tagger for evaluating a convolved uncertainty. 
The photon-beam energy resolution worse than the detector resolution of the tagger 
is caused by the energy spread ($0.11$\%) and angular divergence ($8$ and 
$0.7$~$\mathrm{\mu rad}$ in the horizontal and vertical directions, respectively)
of the stored electron beam, resulting in the fluctuation of recoil-electron momenta 
and angles. If these effects are taken into account as the uncertainty of tagger hit 
positions, a simulated energy resolution reproduces the measured value.

\begin{figure*}[p]
  \centering
  \includegraphics[width=14cm,bb=0 0 567 549]{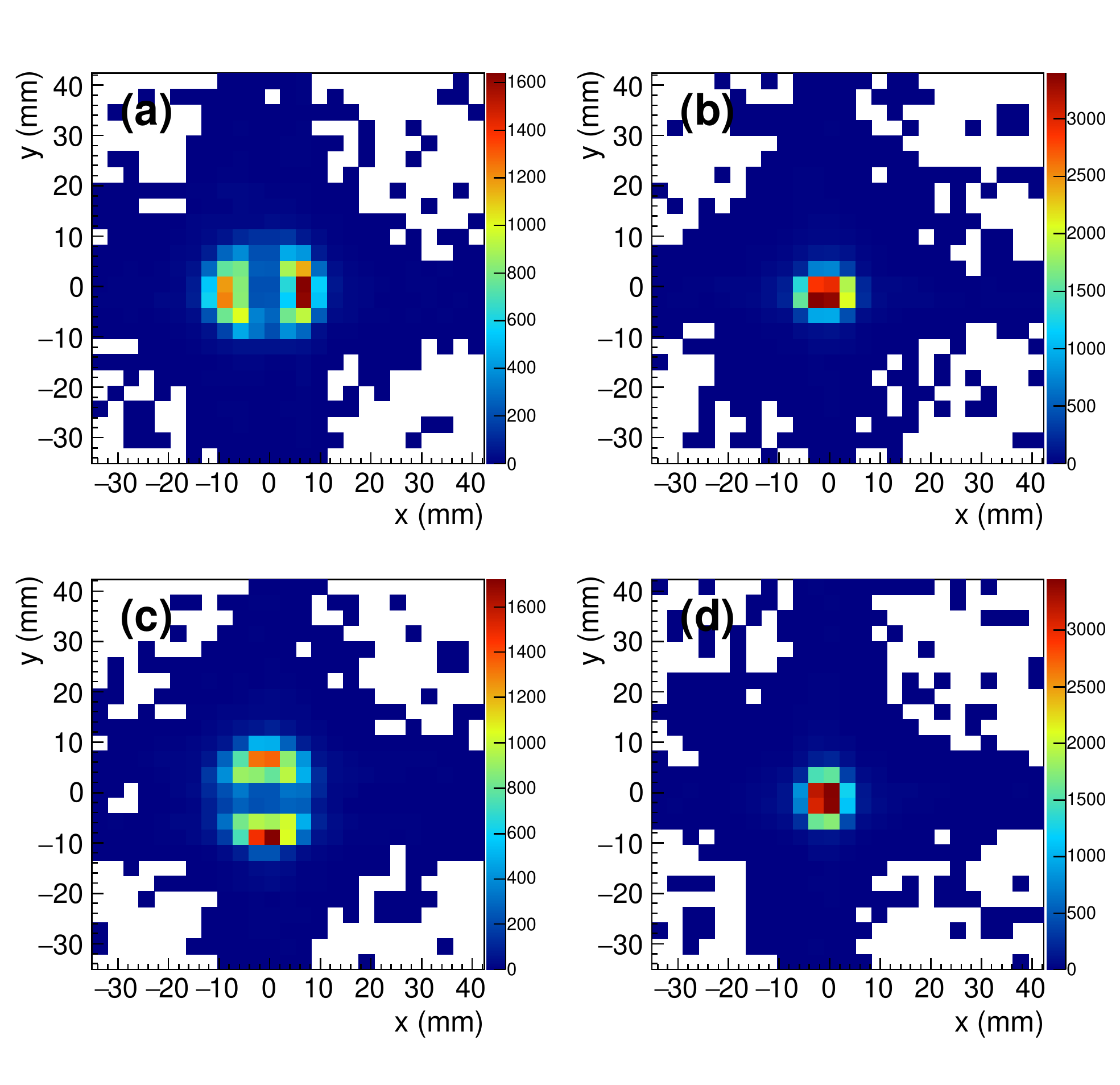}
  \caption{Photon beam profiles when the direction of linear polarization 
           is vertical ((a) and (b)) and horizaontal ((c) and (d)). The left 
           panels ((a) and (c)) are plotted for the photon beam energy range 
           of $0.7$--$1.6$~$\mathrm{GeV}$, while the right panels ((b) and 
           (d)) for that of $1.6$--$2.4$~$\mathrm{GeV}$.}
  \label{fig:beam_profile}
\end{figure*}
Figure~\ref{fig:beam_profile} shows photon beam profiles measured by using a beam 
profile monitor (BPM) in the LEPS2 experimental building \cite{shirai}. The BPM is 
fabricated by arranging 25 scintillating fibers of $3$-$\mathrm{mm}$ square each 
to make a plane sensitve to charged tracks and then by stacking two planes whose 
fiber directions are orthogonal to each other. A signal from each fiber is read out 
by a Multi-Pixel Photon Counter (MPPC) to identify a combination of vertical and 
horizontal fiber channels which an electron and positron pair created at a converter 
have passed through. The $\sigma$ of observed photon beam spread is about 
$3$~$\mathrm{mm}$ for the energy region higher than $1.6$~$\mathrm{GeV}$. It is 
consistent with the beam spread calculated by convolving the electron beam 
divergence with the kinematical angular broadening of laser Compton scattering, 
as shown in Fig.~\ref{fig:beam_divergence}, and taking into account the distance 
of photon beam propagation to the LEPS2 experimental building. 

Because the electron beam divergence of the LEPS2 beamline is much better than that 
of the LEPS beamline, it can be observed that a beam size changes depending on the 
photon beam energy, as expected from the kinematics of laser Compton scattering. 
(Compare Fig.~\ref{fig:beam_profile} (a) and (b), or (c) and (d).) It is also seen 
that the major-axis direction of the beam profile changes by $90$ degrees depending 
on the direction of linear polarization. (Compare Fig.~\ref{fig:beam_profile} (a) 
and (c).) This means that radiation tends to occur at the angles perpendicular to 
the linear polarization direction or the direction of the electric field oscillation. 
This behavior agrees with a simulation result based on the formula by QED \cite{sunwu}.

\section{Summary and future prospects} \label{sec:summary}

The second laser-Compton-scattering facility, called the LEPS2 beamline, has been 
constructed at SPring-8, which is a storage ring of $8$-$\mathrm{GeV}$ electrons.
A linearly polarized photon beam is available in a tagged energy range of 
$1.3$--$2.4$~$\mathrm{GeV}$ by the injection of UV laser light with a wavelength 
of $355$~$\mathrm{nm}$, and hadron photoproduction experiments are already underway. 
The photon beam energy can be determined for each event by measuring the momentum 
of a recoil electron at the tagger, where the lower limit of the tagged energy 
region is lowered to $1.3$~$\mathrm{GeV}$ from that of the LEPS beamline 
($1.5$~$\mathrm{GeV}$). The photon-beam energy resolution is estimated to be 
$12.1$~$\mathrm{MeV}$. Since the electron beam divergence is small at the long 
straight section of the storage ring used by the LEPS2 beamline, the photon beam 
generated by Compton scattering does not spread and can be extracted over a long 
distance. Therefore, the LEPS2 experimental building independent of the building 
housing the storage ring has been constructed at a location about $130$~$\mathrm{m}$ 
away from the Compton scattering point. The construction of the LEPS2 experimental 
building has enabled the installation of two large 
detector systems covering almost all the solid angles. For enhancing the intensity 
of the photon beam reaching the LEPS2 experimental building, the LEPS2 beamline is 
designed to allow the simultaneous injection of four laser beams at a maximum and 
increase the transmittance of the generated photon beam. The internal diameters 
of ultra-high vacuum chambers connected to the storage ring are expanded so that 
multiple laser beams can pass through. The shape of the first mirror used for 
laser injection in a vacuum is designed to avoid direct hits of $X$ rays by 
synchrotron radiation and $\gamma$ rays by laser Compton scattering. The photon 
beam intensity is evaluated by counting the number of tagger trigger signals.
In the LEPS beamline, the tagger trigger counts were largely contaminated by the 
influence of electromagnetic shower produced at the internal structure of the 
vacuum chamber having a window for the tagger. Thus, the corresponding chamber 
in the LEPS2 beamline has been improved so as to reduce such contamination by 
removing a wall structure which is hit by high-momentum recoil electrons. The 
maximum intensity of the photon beam that can be used for hadron photoproduction 
experiments in the LEPS2 experimental building was $2.3 \times 10^6$~$\mathrm{s^{-1}}$. 
By combining the photon beam having high intensity and linear polarization with 
the detector systems having high resolutions and acceptances, we can systematically 
study the nature of hadrons.

In the near future, we will introduce the latest type of pulsed lasers \cite{pulselaser} 
in order to further increase the photon beam intensity. Recently, Spectronics Corp.~has 
released a pulsed UV laser ($355$-$\mathrm{nm}$ wavelength) capable of synchronous output 
by an external signal. It is now possible to inject pulsed laser light at the timing 
synchronized with an electron beam bunch in the storage ring \cite{pulsetest1, pulsetest2}. 
In contrast to the case of continuous wave (CW) or pseudo-CW laser light, the incident 
laser pulse can efficiently collide with electrons at the focal point without wasting 
any laser power. Thus, significant improvement of the photon beam intensity is expected. 
In the LEPS2 beamline, a finite incident angle of laser light is inevitably needed due 
to the simultaneous injection of multiple laser beams. The decrease of a photon beam 
intensity due to the crossing of the electron and laser beam axes around the laser focus 
can be minimized in the case of using a pulsed laser because their scattering positions 
are always fixed at a single point. We have conducted operational tests of the pulsed UV 
laser at the LEPS beamline, and the results will be discussed in a separate paper with 
a detailed description of the method. 

This new laser technology is available not only for the UV wavelength that is currently 
used but also for deep-UV wavelengths that can provide higher energy photon beams. In 
particular, the output powers of pulsed deep-UV lasers have been much improved. For the 
wavelength of $266$~$\mathrm{nm}$, a new pulsed laser developed by Spectronics Corp.~has 
achieved an output power of about $20$~$\mathrm{W}$. This value is an order of magnitude 
higher than the output power of existing deep-UV lasers and rather comparable to that 
of high-power UV lasers. In addition, a pulsed deep-UV laser whose output power is in 
the watt class for a wavelength of $213$~$\mathrm{nm}$ has become a reality. If it is 
used at SPring-8, the maximum energy of a photon beam reaches $3.3$~$\mathrm{GeV}$.

\section*{}
We would like to thank the staff at SPring-8 for supporting the construction and 
comissioning of the LEPS2 beamline and giving excellent experimental conditions. 
We also thank Y.~Ishizawa for supporting the maintenance of our interlock system.
We appreciate the RIKEN for financially supporting the construction of the LEPS2 
experimental building. We are grateful to the technical staff at ELPH, Tohoku
University for supporting the production of electronic circuits for the tagger
and a beam profile monitor. The use of the BL31LEP of SPring-8 (the LEPS2 beamline)
has been approved by the Japan Synchrotron Radiation Institute (JASRI) as a contract 
beamline (Proposal Nos.~BL31LEP/6101 and 6102). This research is supported in part 
by the Ministry of Education, Science, Sports and Culture of Japan.

\end{document}